 \documentclass[10pt,twocolumn,twoside]{IEEEtran}
\usepackage{graphicx} 
\usepackage{amsmath,bm,times} 
\usepackage{amssymb} 
\usepackage{tikz}
\usetikzlibrary{shapes,arrows,backgrounds,fit,positioning}
\usepackage{subfigure}
\usepackage{cite}
\usepackage{soul}
\usepackage{balance}
\usepackage{tabularx}
\usepackage{url}
\allowdisplaybreaks
\usepackage{multirow}
\usepackage{epstopdf}

\newtheorem{assumption}{Assumption}

\newtheorem{definition}{Definition}

\usepackage{algorithm}
\usepackage{algpseudocode}

\title{Multi-Class Stackelberg Games for the Co-Design of Networked Systems}

\author{Julian~Barreiro-Gomez,~\IEEEmembership{Senior Member, IEEE},~and Ye~Wang,~\IEEEmembership{Senior Member, IEEE}
  \thanks{Julian Barreiro-Gomez is with KU Center for Autonomous Robotic Systems, Department of Computer and Information Engineering, Khalifa University, Abu Dhabi 127788, UAE. (E-mail: {\tt\scriptsize julian.barreirogomez@ku.ac.ae})}
  \thanks{Ye Wang is with School of Mathematics and Statistics, The University of Melbourne, Parkville, VIC, 3010, Australia. (E-mail: {\tt\scriptsize ye.wang@unimelb.edu.au})} 
  \thanks{Julian Barreiro-Gomez is profoundly grateful to God for the blessings of health and life, and to Our Lady of Lourdes, whose intercession made this work possible. The work of Julian Barreiro-Gomez is supported by the Khalifa University Faculty Start-Up (FSU) Grant 8474000774. Ye Wang acknowledges support from the Australian Research Council through the Discovery Early Career Researcher Award (DE220100609).}
}

\IEEEoverridecommandlockouts

\usepackage{algorithm}
\usepackage{algpseudocode}
\usepackage[mathscr]{euscript}

\begin{document}

 \maketitle

\begin{abstract}
We investigate a co-design problem, which entails the simultaneous optimization of both system infrastructure and control strategies, within a game-theoretical framework. To this end, we formulate the co-design problem as a two-layer hierarchical strategic interaction. At the upper layer, a leader (or multiple leaders) determines system design parameters, while at the lower layer, a follower (or multiple followers) optimizes the control strategy. To capture this hierarchy, we propose four novel classes of Stackelberg games that integrate diverse strategic behaviors, including combinations of cooperative and non-cooperative interactions across two different layers. Notably, the leaders' interactions are represented using a normal-form game, whereas the followers' interactions are modeled by difference games (dynamic games in discrete time). These distinct game structures result in a Stackelberg game formulation that accommodates different game types per layer, and/or supports heterogeneous strategic behaviors involving cooperation and non-cooperation simultaneously. Learning algorithms using the best-response dynamics are used to solve the game problems when considering a discrete strategic space for the leaders. The effectiveness of the proposed approach is demonstrated through an application to the co-design of the Barcelona drinking water network.
\end{abstract}


\vspace{-2mm}
\section{Introduction}

\IEEEPARstart{G}{ame} theory provides a rigorous mathematical framework for analyzing and modeling strategic interactions among rational decision-makers. Its versatility has led to widespread applications in engineering, including power systems \cite{churkin2021review}, robotic coordination \cite{pradhan2021game}, 
and cyber-physical systems \cite{tushar2023survey}, among others. The literature identifies several classes of games that capture distinct interaction dynamics, such as cooperation, competition, altruism, and co-opetition.

Several classes of games have been extensively studied and reported in the literature, each tailored to represent specific interactive scenarios. One of the most fundamental and widely recognized solution concepts in non-cooperative game theory is the Nash equilibrium, which characterizes a stable outcome wherein no agent has an incentive to unilaterally deviate from its current strategy \cite{nash2024non}. Beyond this, the game theory community has developed a variety of other frameworks to capture more nuanced interaction structures. \emph{Stackelberg game} is a hierarchical decision-making process, where one player (the leader) moves first, and the others (the followers) react accordingly \cite{Stackelberg_1948}, \cite[Chapter 7]{BookBaTe}. Coalitional games have been developed to analyze the formation of alliances, compute power indices, and assess the influence of individual participants in cooperative environments \cite{Holler2001}. Similarly, cooperative games explore settings in which agents seek to maximize a collective payoff, while Berge games are used to formalize altruistic behaviors in strategic contexts \cite{Berge}. The concept of co-opetition, which blends elements of cooperation and competition, has been formalized to describe many real-world scenarios where agents must balance conflicting objectives \cite{Coopetitive}, \cite[Chapter 5]{BookBaTe}. Given this diversity of models, it is well-established that game theory provides a powerful theoretical tool for designing and analyzing decentralized control strategies in complex systems.

Modern networked systems are inherently complex and large-scale, comprising multiple interconnected subsystems that must coordinate to achieve overall system efficiency and reliability \cite{zhang2024,liu2020coordinated}. Examples of such systems include water distribution networks \cite{wang2017non,arastou2025optimization}, power grids \cite{Yu_2025}, supply chains \cite{qian2022integrated}, and transportation systems \cite{Huzaefa2023}. The design and control of these systems pose significant challenges, particularly in industrial applications where infrastructure is often developed based on conservative estimates, with advanced control strategies introduced only afterward. This sequential design–then–control approach leads to a suboptimal solution, as the control strategy is constrained by the limitations of a predetermined system design. To address this limitation, we focus on \emph{co-design problem}, an approach that seeks to simultaneously optimize both the system infrastructure and the associated control strategy \cite{wang2025tcst}. By jointly considering these two aspects, the co-design framework has the potential to achieve globally optimal solutions with respect to the overall system objectives, thereby overcoming the inherent inefficiencies of sequential design.

In this paper, we study the co-design problem based on Stackelberg game. The Stackelberg game framework, originally proposed by Heinrich von Stackelberg in \cite{Stackelberg_1948}, models hierarchical decision-making structures involving a leader and one or more followers. The use of leader-follower schemes has been successfully applied to solve several engineering problems, e.g., formation control \cite{barreiro2021distributed,wang2021path}, crowd dynamics modeling \cite{BarreiroMasmoudi_23}, hierarchical scheme for price dynamics analysis under the framework of mean-field-type games \cite{hierarchical2022price}, consensus for nonlinear multi-agent systems \cite{Zhang_2025}. %
In its original formulation, Stackelberg analyzed the interactions between firms with asymmetric decision-making, where the leader commits to a strategy first, and the followers respond optimally to that strategy. This hierarchical, bi-level decision-making framework is well-suited for addressing the co-design problem, as it mirrors the natural structure of infrastructure design (leader) followed by operational control (follower). Accordingly, we adopt a Stackelberg game formulation to model the inter-dependencies between system design and control, thereby providing a structured methodology for deriving optimal co-design solutions.

Furthermore, we extend this modeling framework by considering the possibility of strategic interactions within each layer of the hierarchical system. From a global perspective, we conceptualize the problem as consisting of two interacting layers: the designer of the system and the operator responsible for its ongoing functionality. Within the design layer, we may encounter multiple decision-makers with potentially conflicting objectives, or alternatively, a centralized designer solving an optimization problem on behalf of the entire system. Similarly, on the operational side, we may face either centralized or decentralized control strategies, depending on the system characteristics and operational constraints. This layered and strategic view provides a flexible and comprehensive framework for modeling complex co-design problems in large-scale networked systems.

The main contribution of this paper is to propose a two-layer hierarchical game or a Stackelberg-like game problem that combines at each layer different classes of games. For the leaders' layer, we have a normal form non-cooperative game with a finite number of strategies and whose utility functions depend on both the leaders' and followers' strategic selection. Hence, at this leaders' layer, we may also consider cooperation, in which we perform optimization for the design parameters of the system. Regarding the followers' layer, we introduce a dynamic game, more precisely, a non-cooperative difference game problem. Similarly, we may also consider a cooperative game at the followers' layer, which corresponds to a standard optimal control problem \cite{kirk2004optimal}. We highlight that the interaction between the layers in the hierarchical scheme is different from others reported in the literature. The hierarchical scheme of coupling is mainly given by the fact that the leader defines the feasible set of strategies for followers.

We show that the proposed Stackelberg game, which combines leader normal form games with discrete strategic sets and follower difference games (dynamic games), can be used for solving the co-design control and system problem. The four Stackelberg game classes that we study in this paper are:
\begin{itemize}
    \item Non-cooperative leaders and followers,
    \item Cooperative leaders and followers,
    \item Non-cooperative leaders and cooperative followers,
    \item Cooperative leaders and non-cooperative followers,
\end{itemize}
which will be formally introduced and explained later on throughout the paper. 

The remainder of this paper is organized as follows. Section \ref{sec:problem_statement} presents the preliminaries comprising the game settings for both the leader and follower decision-makers, game-theoretical solution concepts, and the price of anarchy in this co-design problem context. Section \ref{sec:stackelberg} introduces the proposed four Stackelberg game classes together with their corresponding Stackelberg equilibria. Section \ref{sec:case_study} presents the networked system application we use to illustrate the contributions of this paper. In Section \ref{sec:results}, the results are presented and discussions are developed to compare the different Stackelberg game classes. Finally, concluding remarks and future directions are summarized in Section \ref{sec:conclusions}. %

\vspace{-3mm}
\section{Problem Statement}\label{sec:problem_statement}

In this paper, we consider a networked engineering system and aim to analyze two main problems using game-theoretical tools. On one hand, we analyze the design of the networked system led by a leader's layer. At this design stage, some key parameters for the system are determined. On the other hand, we analyze the control design to operate the system led by a followers' layer. At this design stage, optimal control policies are defined to operate the system subject to the established design parameters by the leaders.
There are two classes of decision-makers, i.e., leaders and followers. The set of $L \in \mathbb{N}$ leaders is given by $\mathcal{L} := \{1,\dots,L\} $, and the set of $M \in \mathbb{N}$ followers is given by $\mathcal{M}:=\{1,\dots,M\}$. The leader decision-makers set takes care of the system design and parameter setting, whereas the second decision-makers set is in charge of designing the optimal control to operate the system, adhering to the imposed rules of the leader's set. The leaders interact with each other using a static normal form game, and their decisions are coupled with the followers' interactions. The followers interact with each other by means of a difference game whose settings are leader-strategic-dependent. 

In the following, we explain the settings for each game layer, i.e., for the leaders and followers, and then we formally introduce the corresponding game problems together with the different game-theoretical solution concepts we are interested in. Let us start with the followers' strategic interaction settings.

\vspace{-3mm}
\subsection{Followers Strategic Interaction and Settings}

The followers interact on a difference game problem within a discrete-time interval $[0..T] := [0,T] \cap \mathbb{Z}_+$, with $T\in \mathbb{N}$, involving a dynamical system given as follows:
\begin{subequations}\label{eq:dynamical_system}
    \begin{align}
        x_{k+1} &= f(x_{k},\{u_{i,k}\}_{i \in \mathcal{M}}),~\forall~k \in [0..(T-1)],\\
        x_0 &\in \mathbb{X}(\cdot)~\text{given}, 
    \end{align}
\end{subequations}
where $x \in \mathbb{X}(\cdot) \subset \mathbb{R}^{n_x}$ denotes the system states. We highlight that the feasible set for the system states depends on a given parameters, which comes from the leader's strategic selection, i.e., $\mathbb{X}(\{\text{leader strategies}\})$. The strategic selection of the $i-$th follower decision-maker is denoted by $u_{i} \in \mathbb{U}_i \subset \mathbb{R}^{n_{u_i}}$. Let $$\mathcal{U}_i:= \{ \mathbf{u}_{i} := u_{i,0:T-1} : u_{i,k} \in \mathbb{U}_i,~\forall~k \in [0..(T-1)]\} $$ denote the set of admissible controls of the $i$-th decision-maker. Also, we denote $\mathbf{u}_{-i}$ as the strategic selection sequences along the time horizon for all the decision-makers different from $i$, i.e., 
\begin{align}
    \mathbf{u}_{-i} &:= (\mathbf{u}_{1}, \dots, \mathbf{u}_{i-1}, \mathbf{u}_{i+1},\dots,\mathbf{u}_{M}) \in \mathcal{U} = \hspace{-0.2cm}\prod_{j \in \mathcal{M}\setminus\{i\}} \mathcal{U}_j,
\end{align}
where $\mathbf{u} := (\mathbf{u}_{1}, \dots, \mathbf{u}_{M})$ is the strategic profile or the joint strategic actions for the $M$ follower decision-makers. 

\subsection{Followers Non-Cooperative Behavior}

Each decision-maker seeks to minimize its own cost functional given by $V_i(x_0,\mathbf{u}_{i},\mathbf{u}_{-i})$, and the difference game problem $\mathscr{P}^{\mathrm{NC}}_{\mathrm{follower}}$ is expressed as follows:
\begin{align}
    \label{eq:follower_problem}
    &\mathscr{P}^{\mathrm{NC}}_{\mathrm{follower}}:~ \forall~i\in \mathcal{M},\\
    &\min_{\mathbf{u}_{i} \in \mathcal{U}_i} V_i(x_0,\mathbf{u}_{i},\mathbf{u}_{-i})~\text{s.~t.}~ 
    \begin{cases}
        \eqref{eq:dynamical_system}, \\
        x_k \in \mathbb{X}(\{\text{leader strategies}\}),\\ 
        \text{for all}~ k\in[0..T],
    \end{cases} \notag
\end{align}
where each cost $V_i$ is assumed to be continuous, convex and coercive in $\mathbf{u}_{i}$, for all $i \in \mathcal{M}$, such that the optimization problem is well-defined. Given that the follower layer game problem depends on the decisions made at the leader level, we make the following assumption.

\begin{assumption}\label{assump:1}
    The optimization problem \eqref{eq:follower_problem} is feasible for every leader strategic profile $\mathbf{a} \in \mathcal{A}$. \hfill $\square$   
\end{assumption}

Once the follower's game problem is defined, we introduce the equilibrium concept by using the best-response strategies. 

\begin{definition}[Best-response strategies among followers]\label{def:br_followers}
    A feasible control strategy $\mathbf{u}_{i} \in \mathcal{U}_i$ is a best response strategy if it solves the Problem in \eqref{eq:follower_problem}, given the strategic selection by other decision-makers $\mathcal{M}\setminus\{i\}$ given by $\mathbf{u}_{-i}$ and the strategic selection of the leader(s). The set of best-response strategies for the $i$-th decision-maker is denoted by $\mathbb{BR}^\mathcal{M}_i(\mathbf{u}_{-i},\{\text{leader strategies}\})$. \hfill $\square$   
\end{definition}

Then, we use the best-response strategy concept to introduce the definition of game-theoretical equilibrium below.
\begin{definition}[Nash equilibrium among followers]
    \label{def:Nash_followers}
    A strategic profile $\mathbf{u}^* = (\mathbf{u}^*_{1},\dots,\mathbf{u}^*_{M}) \in \mathcal{U}$ is a Nash equilibrium if all the strategies are best-response strategies against each other given the strategic selection of the leader(s), i.e., $\mathbf{u}^*_{i} \in \mathbb{BR}^\mathcal{M}_i(\mathbf{u}^*_{-i},\{\text{leader strategies}\})$, for all $i \in \mathcal{M}$. The set of Nash equilibria for the followers' strategic interaction is denoted by $\mathbb{NE}_{\mathrm{follower}}$. \hfill $\square$
\end{definition}

The following definition introduces the concept of $\varepsilon$ equilibrium, in which we can finish the solution seeking.
\begin{definition}[$\varepsilon$-Nash equilibrium among followers]
\label{def:epsilon_nash}
    A strategic profile $\mathbf{u}^* = (\mathbf{u}^*_{1},\dots,\mathbf{u}^*_{M}) \in \mathcal{U}$ is a $\varepsilon$-Nash equilibrium if
    \begin{align}
        V_i(x_0,\mathbf{u}_{i}^*,\mathbf{u}_{-i}^*) \leq V_i(x_0,\mathbf{u}_{i},\mathbf{u}_{-i}^*) + \varepsilon,
    \end{align}
    for all $\mathbf{u}_i \in \mathcal{U}_i$, and $i \in \mathcal{M}$. 
    The set of $\varepsilon$-Nash equilibria is denoted by $\varepsilon$-$\mathbb{NE}_{\mathrm{follower}}$. \hfill $\square$
\end{definition}

\begin{definition}[Best-response of non-cooperative followers strategies against leaders' strategies]
    A feasible strategic follower profile $\mathbf{u}^* \in \mathcal{U}$ is a non-cooperative best response against the strategic selection of the leader strategy if $\mathbf{u}^*$ is a Nash equilibrium. We denote by $\mathbb{BR}^{\mathrm{NC}}_\mathcal{M}(\{\text{leader strategies}\})$ the set of non-cooperative best response strategies given the strategic selection by the leader, i.e., $\mathbf{u}^* \in \mathbb{BR}^{\mathrm{NC}}_\mathcal{M}(\{\text{leader strategies}\})$.  \hfill $\square$
\end{definition}

We would also like to study the case in which the followers interact in a cooperative framework. This strategic behavior is formally presented in the next subsection.

\subsection{Followers Cooperative Behavior}

The followers can interact in a cooperative game by jointly optimizing a common cost functional. The cooperative game problem, which corresponds to a standard optimal control problem, is formulated as follows:
\begin{align}
\label{eq:follower_problem_coop}
    &\mathscr{P}^{\mathrm{C}}_{\mathrm{follower}}: \\
    &\min_{\mathbf{u} \in \mathcal{U}} \sum_{i \in \mathcal{M}} V_i(x_0,\mathbf{u}_{i},\mathbf{u}_{-i}), ~\text{s.~t.}
    \begin{cases}
        \eqref{eq:dynamical_system},\\ x_k \in \mathbb{X}(\{\text{leader strategies}\}),\\ \text{for all}~ k\in[0..T].
    \end{cases}\notag
\end{align}
For simplicity in the notation, let us consider
\begin{align*}
    V(x_0,\mathbf{u}) =
       \sum_{i \in \mathcal{M}} V_i(x_0,\mathbf{u}_{i},\mathbf{u}_{-i}).
\end{align*}

\vspace{-3mm}
Similar to Assumption \ref{assump:1} stated in the non-cooperative game problem settings, the following assumption is made for the cooperative case.

\begin{assumption}
\label{assump:2}
    The optimization problem \eqref{eq:follower_problem_coop} is feasible for every leader strategic profile $\mathbf{a} \in \mathcal{A}$. \hfill $\square$
\end{assumption}

Next, we introduce the best action that a follower can take, in a cooperative way, against the decisions made by the leader, in Definition \ref{def:br_follow_against_leaders}. Note that this is different from Definition~\ref{def:br_followers}.

\begin{definition}[Best-response of cooperative followers strategies against leaders' strategies]
\label{def:br_follow_against_leaders}
    A feasible strategic follower profile $\tilde{\mathbf{u}}^* \in \mathcal{U}$ is a cooperative best response against the strategic selection of the leader strategy if it solves the Problem in \eqref{eq:follower_problem_coop}, i.e., 
\begin{align}
    \tilde{\mathbf{u}}^*(\{\text{leader strategies}\}&) \in \arg\min_{\mathbf{u} \in \mathcal{U}} \sum_{i \in \mathcal{M}} V_i(x_0,\mathbf{u}_{i},\mathbf{u}_{-i}), \notag\\
    \text{s.~t.} &
    \begin{cases}
        \eqref{eq:dynamical_system},\\ x_k \in \mathbb{X}(\{\text{leader strategies}\}), \\
        \text{for all}~ k\in[0..T],
    \end{cases}  
\end{align}

\vspace{-3mm}
We denote $\mathbb{BR}^{\mathrm{C}}_\mathcal{M}(\{\text{leader strategies}\})$ as the set of cooperative best response strategies given the strategic selection by the leaders. \hfill $\square$
\end{definition}

\vspace{-4mm}
\subsection{Leaders Strategic Interaction and Settings}
The leaders action set is $\{A_{1},\dots,A_{L}\}$ and each set contains $N \in \mathbb{N}$ actions. %
The $j-$th leader chooses an action $a_j \in A_j$. A strategic profile is given by the joint selection of strategies along all the players, i.e.,
\begin{align*}
    \mathbf{a} = (a_1,\dots,a_L) \in \mathcal{A} = \prod_{j \in \mathcal{L}} A_j.
\end{align*}

\vspace{-2mm}
Each leader decision-maker $j \in \mathcal{L}$ has an associated cost function, denoted by $J_{j}(\mathbf{a},\cdot) = J_{j}(a_{j},a_{-j},\cdot)$, for all $j \in \mathcal{L}$, to be minimized. Note that the cost functional of the leader decision-makers depends on the whole leader strategic profile and other terms coming from the decisions made by the followers, i.e., $J_{j}(a_{j},a_{-j},\{\text{follower strategies}\})$. 

\vspace{-4mm}
\subsection{Leaders Non-Cooperative Behavior}

To achieve a non-cooperative behavior, each leader makes decisions by solving the following optimization problem:
\begin{align}
\label{eq:leader_problem}
    &\mathscr{P}^{\mathrm{NC}}_{\mathrm{leader}}:~ \forall~j\in \mathcal{L}, \notag\\ 
    &\min_{a_j \in A_j} J_{j}(a_{j},a_{-j},\{\text{follower strategies}\}),
\end{align}
where the cost functional $J_j$ is assumed to be continuous, convex, and coercive in the strategies $a_j$, for all $j\in \mathcal{L}$, such that the optimization problem is well defined. We next introduce the best decision made by a leader depends on the decisions made by all the other leaders.

\begin{definition}[Best-response strategies among leaders]    
\label{def:br_leaders}
    A strategy $a_j \in A_j$ is a best response strategy if for a given $a_{-j}$ that is the optimal strategy that minimizes the cost of the $j-$th leader, i.e., if it solves the problem in \eqref{eq:leader_problem}. The set of best response strategies for the $j-$th leader is denoted by $\mathbb{BR}^\mathcal{L}_j(a_{-j},\{\text{follower strategies}\})$.  \hfill $\square$
\end{definition}

We introduce the game-theory equilibrium concept in Definition \ref{def:Nash_leaders} by using the best-response strategy in Definition  \ref{def:br_leaders}.

\begin{definition}[Nash equilibrium among leaders]   
\label{def:Nash_leaders}
    A feasible strategic profile $\mathbf{a}^* := (a_1^*,\dots,a_L^*) \in \mathcal{A}$ is a Nash equilibrium if no leader has incentives to unilaterally change its strategy,
    \begin{align}
        J_{j}(a^*_{j},a^*_{-j},&\{\text{follower strategies}\}) \notag\\
        &\leq J_{j}(a_{j},a^*_{-j},\{\text{follower strategies}\}),
    \end{align}
    for all $a_j \in A_j$, and $j \in \mathcal{L}$. Alternatively, the strategic profile $\mathbf{a}^* := (a_1^*,\dots,a_L^*)$ is a Nash equilibrium if all strategies are best-response strategies against each other, i.e., $a_j^* \in \mathbb{BR}^\mathcal{L}_j(a_{-j},\{\text{follower strategies}\})$, for all $j \in \mathcal{L}$. The set of Nash equilibria for the leaders strategic interaction is denoted by $\mathbb{NE}_{\mathrm{leader}}$.  \hfill $\square$
\end{definition}

There are two important aspects to mention related to the game settings for the leaders' problem. On one hand, note that the strategic set is considered to be discrete, i.e., a finite number of possible decisions that the leaders can make. On the other hand, we are only evaluating the pure strategies for the leaders. Therefore, we need to set the following assumption.

\begin{assumption}
\label{assump:3}
    The normal form game problem presented in \eqref{eq:leader_problem} admits a Nash equilibrium in pure strategies introduced in Definition \ref{def:Nash_leaders}. \hfill $\square$
\end{assumption}

Note that the relaxation of Assumption \ref{assump:3} is challenging as it implies obtaining an expression of the optimal solution for the followers in terms of the leader's decision. The computation of this optimal representation, for the followers in terms of the leaders, is not that involved when the leaders' actions affect either the dynamical system or the cost for the followers. In contrast, in this problem setting, the leaders' decisions affect the feasible set of the system states. 

In Section \ref{sec:stackelberg}, we will present how to approximate this solution by using an algorithm, in which we use an $\epsilon$ equilibrium condition that is introduced next in Definition \ref{def:epsilon_nash_2}.

\begin{definition}[$\varepsilon$-Nash equilibrium among leaders]    
\label{def:epsilon_nash_2}
A feasible strategic profile $\mathbf{a}^* := (a_1^*,\dots,a_L^*) \in \mathcal{A}$ is an $\varepsilon$-Nash equilibrium if
\begin{align}
    J_{j}(a^*_{j},a^*_{-j},&\{\text{follower strategies}\}) \notag\\
    &\leq J_{j}(a_{j},a^*_{-j},\{\text{follower strategies}\}) + \varepsilon,
\end{align}
for all $a_j \in A_j$, and $j \in \mathcal{L}$. 
The set of $\varepsilon$-Nash equilibria is denoted by $\varepsilon$-$\mathbb{NE}_{\mathrm{leader}}$.  \hfill $\square$
\end{definition}

To complete all the possible combinations in the interactions among leaders and followers, we also consider the case in which the leaders behave in a cooperative way, as introduced in the following subsection.

\subsection{Leaders Cooperative Behavior}

The leaders can cooperate by jointly optimizing the cost functionals related to the design stage. Therefore, the cooperative leader problem is as follows:
\begin{align}
\label{eq:leader_problem_coop}
    &\mathscr{P}^{\mathrm{C}}_{\mathrm{leader}}:~  \min_{\mathbf{a} \in \mathcal{A}} \sum_{j \in \mathcal{M}} J_{j}(a_{j},a_{-j},\{\text{follower strategies}\}).
\end{align}
For notation simplicity, let $J(\mathbf{a},\{\text{follower strategies}\})$ be as
\begin{align*}
J(\mathbf{a},\cdot) = 
    \sum_{j \in \mathcal{M}} J_{j}(a_{j},a_{-j},\{\text{follower strategies}\}).
\end{align*}

\begin{definition}[Cooperative solution for the leaders]
    A strategic profile $\tilde{\mathbf{a}}^* = (a_1^*,\dots,a_{L}^*) \in \mathcal{A}$ is a cooperative solution if it solves \eqref{eq:leader_problem_coop}, i.e., $$\tilde{\mathbf{a}}^* \in \arg\min_{\mathbf{a} \in \mathcal{A}} \sum_{j \in \mathcal{M}} J_{j}(a_{j},a_{-j},\{\text{follower strategies}\}).\; \hfill \square$$
\end{definition}

There are some advantages to solving the problem following a non-cooperative approach for networked systems, e.g., when solving the non-cooperative game problem for the followers, the structure of the controller can be interpreted as a non-centralized controller. Whereas when the followers cooperate, this architecture can be seen as a centralized controller. Therefore, we are interested in judging and measuring how different the non-cooperative and cooperative solutions are. To this end, we use the price of anarchy concept.

\vspace{-4mm}
\subsection{Price of Anarchy for Leaders and Followers}

The price of anarchy is a key performance indicator that allows measuring how optimal a Nash equilibrium is with respect to the best socially optimal combination \cite{PoA_2024}. In this work, we take the cost for a Nash equilibrium in comparison to the cost when all the decision-makers cooperate with one another to pursue a social optimum. Indeed, notice that, according to the Definition \ref{def:Nash_followers} and Definition \ref{def:Nash_leaders}, there must be multiple Nash equilibrium points. If there are multiple Nash equilibria, then the price of anarchy is computed using the worst equilibrium in terms of its corresponding cost. Suppose you have the set of Nash equilibria for the leaders as $\mathbb{NE}_\mathrm{leader}$ and for the followers $\mathbb{NE}_\mathrm{follower}$. Then, the corresponding prices of anarchy are defined as follows:
\begin{subequations}
    \begin{align}
        \mathrm{PoA}_{\mathrm{leader}} &= \dfrac{\max\limits_{\mathbf{a} \in \mathbb{NE}_{\mathrm{leader}}} J(\mathbf{a},\mathbf{u})}{\min\limits_{a \in \mathcal{A}} J(\mathbf{a},\mathbf{u})} \geq 1, \label{eq:poa_leaders}\\
    %
        \mathrm{PoA}_{\mathrm{follower}} &= \dfrac{\max\limits_{\mathbf{u} \in \mathbb{NE}_{\mathrm{follower}}} V(x_0,\mathbf{u})}{\min\limits_{u \in \mathcal{U}} V(x_0,\mathbf{u})} \geq 1. \label{eq:poa_followers}
    \end{align}
\end{subequations}

Note that the denominators in both expressions for the price of anarchy are computed by solving the cooperative game problem presented in Problem \eqref{eq:leader_problem_coop} and Problem \eqref{eq:follower_problem_coop}. Therefore, the denominators become $J(\tilde{\mathbf{a}}^*,\{\text{follower strategies}\})$ and $V(x_0,\tilde{\mathbf{u}}^*)$, respectively. In addition, let us assume that, as we will see in the numerical example we present in this paper, there is a unique Nash equilibrium for each of the games across the layers, i.e. ${\mathbf{a}}^*$ and ${\mathbf{u}}^*$. Then, the prices of anarchy can be written as
\begin{align*}
    \mathrm{PoA}_{\mathrm{leader}} &= \dfrac{J(\mathbf{a}^*,\mathbf{u})}{J(\tilde{\mathbf{a}}^*,\mathbf{u})}, &
    %
    \mathrm{PoA}_{\mathrm{follower}} &= \dfrac{V(x_0,\mathbf{u}^*)}{V(x_0,\tilde{\mathbf{u}}^*)},
\end{align*}
respectively. A lower price of anarchy corresponds to an enhancement in the worst scenario for the equilibrium performance, and $\mathrm{PoA}_{\mathrm{leader}} = 1$ or $\mathrm{PoA}_{\mathrm{follower}}=1$ imply that the Nash equilibria are optimal. In other words, when having a unitary price of anarchy, the non-centralized design by the leaders using normal form games, or the non-centralized design of the control law by the dynamic games of followers, are as optimal as those obtained when they jointly optimize in the framework of cooperation.

\begin{table}[t]
    \centering
    \caption{Different classes of Stackelberg games}
    \label{tab:stackelberg_classes}
    \resizebox{\columnwidth}{!}{
    \begin{tabular}{cc|cc}
        \hline
       \textbf{Stackelberg} & \textbf{Strategic}  & \textbf{Leaders Interaction} & \textbf{Followers Interaction}  \\
       \textbf{Game} & \textbf{Profile}  & \textbf{Normal Form Game} & \textbf{Difference Game}  \\
       \hline
       Class I & $(\mathbf{a}^*,\mathbf{u}^*)$ & Non-cooperative & Non-cooperative \\
       Class II & $(\tilde{\mathbf{a}}^*,\tilde{\mathbf{u}}^*)$ & Cooperative & Cooperative \\ 
       Class III & $(\mathbf{a}^*,\tilde{\mathbf{u}}^*)$ & Non-cooperative &  Cooperative \\ 
       Class IV & $(\tilde{\mathbf{a}}^*,\mathbf{u}^*)$ & Cooperative & Non-cooperative \\ 
       \hline
    \end{tabular}
    }
\end{table}

\vspace{-3mm}
\section{Stackelberg equilibrium for \\the co-design problem}
\label{sec:stackelberg}

\begin{figure}[t]
    \centering
    \includegraphics[width = 0.8\hsize]{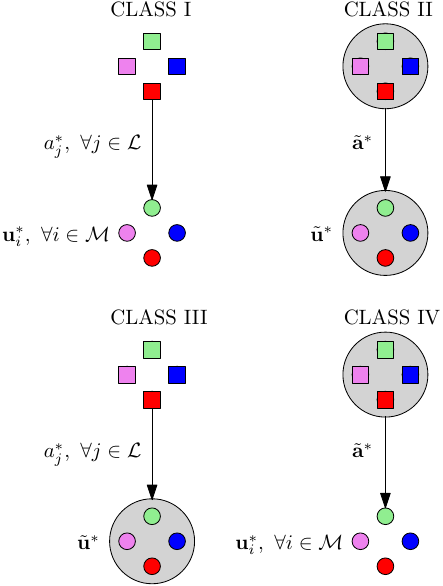}
    \caption{The hierarchical strategic interactions illustrating four Stackelberg game classes (Squares represent multiple leaders and circles represent multiple followers).}
    \label{fig:classes}
\end{figure}

As we can consider multiple strategic interactions at each layer of the hierarchical Stackelberg scheme, we have multiple classes for the Stackelberg game as summarized in Table \ref{tab:stackelberg_classes}. Each class is also presented in Fig.~\ref{fig:classes}. It is worth highlighting that, unlike other works employing Stackelberg games, where strategic coupling between the leader and the follower typically occurs through either the dynamics of the system or cost/utility functional (see \cite{BookBaTe,Jingrui2023,HUANG20202237}), the formulation considered in this work focuses on a different type of interaction. Specifically, we analyze the case in which the leader’s decision affects the feasible set of the optimization problem of the followers. This distinction allows us to capture a broader class of hierarchical decision-making scenarios, particularly relevant for co-design problems where infrastructure choices constrain the subsequent control strategies.

\vspace{-5mm}
\subsection{Definition of Stackelberg Equilibrium Classes}

This subsection introduces the possible strategic interactions at each layer, which establish a Stackelberg game class. Let us consider the case in which the leaders $\mathcal{L}$ do not cooperate, i.e., there are $L$ designer entities making decisions following an independent interest. Also, assume that once the design is determined, the followers $\mathcal{M}$ design a control action independently, i.e., there are $M$ entities deciding control actions following different non-cooperative costs. The Stackelberg equilibrium corresponding to this scenario is formally presented in Definition \ref{def:class_I} below (see Fig. \ref{fig:classes}).

\begin{definition}[Class I: Stackelberg equilibrium with non-cooperative leaders and non-cooperative followers] 
\label{def:class_I}
    A strategic leader-follower profile $(\mathbf{a}^*,\mathbf{u}^*) \in \mathcal{A} \times \mathcal{U}$ is a Stackelberg equilibrium for non-cooperative leaders and followers if
    \begin{subequations}
        \begin{align}
        &\forall~i \in \mathcal{M}:~\mathbf{u}_i^* \in \mathbb{BR}^\mathcal{M}_i(\mathbf{u}_{-i}^*,\mathbf{a}^*),\\
        &\forall~j \in \mathcal{L}: \\
        &a_j^* \in \arg\min_{a_j \in A_j} J_{j}\left(a_{j},a_{-j}^*,\mathbf{u}^* \bigg|
        \begin{array}{l}
              \mathbf{u}_{i}^* \in \mathbb{BR}^\mathcal{M}_i(\mathbf{u}_{-i}^*,a_{j},a_{-j}^*)\\
              \forall i\in\mathcal{M}
        \end{array}
        \hspace{-0.2cm}\right). \notag
    \end{align}
    \end{subequations}
    Note that the leader designs some parameters knowing that the follower is going to optimize based on them. \hfill $\square$
\end{definition}

Let us now assume that all the leaders $\mathcal{L}$ coordinate with each other and agree to jointly perform the design of the system. This cooperation leads to an optimization that is jointly performed by the $L$ leaders. Then, once the system design is established, the followers $\mathcal{M}$ cooperatively design the control actions to operate the system. This cooperative followers' behavior leads to a centralized control design. Definition~\ref{def:class_II} formally presents the Stackelberg equilibrium under this scenario (see Fig. \ref{fig:classes}).

\begin{definition}[Class II: Stackelberg equilibrium with cooperative leaders and cooperative followers] 
\label{def:class_II}
    A strategic leader-follower profile $(\tilde{\mathbf{a}}^*,\tilde{\mathbf{u}}^*) \in \mathcal{A} \times \mathcal{U}$ is a Stackelberg equilibrium for cooperative leaders and followers if
    \begin{subequations}
        \begin{align}
        \tilde{\mathbf{u}}^* &\in \mathbb{BR}^\mathcal{M}(\tilde{\mathbf{a}}^*),\\
        \tilde{\mathbf{a}}^* &\in \arg\min_{\mathbf{a} \in \mathcal{A}} \sum_{j \in \mathcal{M}} J_{j}(\mathbf{a},\tilde{\mathbf{u}}^*|\tilde{\mathbf{u}}^*\in \mathbb{BR}^\mathcal{M}(\tilde{\mathbf{a}})).
    \end{align}
    \end{subequations}
    In this case, all the leaders jointly design parameters against which the followers will optimally respond, also jointly by means of cooperation. \hfill $\square$
\end{definition}

The aforementioned Stackelberg game classes have considered a homogeneous behavior for both the leader and followers, i.e., both layers cooperating or both layers acting independently in a non-cooperative framework. However, layers may exhibit heterogeneous strategic behavior. First, let us consider the case in which the leaders $\mathcal{L}$ do not cooperate for the system design, whereas the followers $\mathcal{M}$ cooperatively react to the system design by solving jointly a centralized optimal control problem. The emerging Stackelberg equilibrium for this scenario is formally presented next in Definition \ref{def:class_III}.

\begin{definition}[Class III: Stackelberg equilibrium with non-cooperative leaders and cooperative followers] 
\label{def:class_III}
    A strategic profile $(\mathbf{a}^*,\mathbf{u}^*) \in \mathcal{A} \times \mathcal{U}$ is a Stackelberg equilibrium for non-cooperative leaders and cooperative followers if
        \begin{align*}
        &\forall~i \in \mathcal{M}:~\mathbf{u}_i^* \in \mathbb{BR}^\mathcal{M}_i(\mathbf{u}_{-i}^*,\tilde{\mathbf{a}}^*),\\
        &\tilde{\mathbf{a}}^* \in \arg\min_{\mathbf{a} \in \mathcal{A}} \sum_{j \in \mathcal{M}} J_{j} \left(\mathbf{a},\mathbf{u}^*|\mathbf{u}_i^*\in \mathbb{BR}^\mathcal{M}_i(\mathbf{u}_{-i}^*,\tilde{\mathbf{a}})~\forall i\in\mathcal{M} \right).
    \end{align*}
    In this case, all the leaders non-cooperatively design the parameters against which the followers will jointly and optimally respond by means of cooperation. \hfill $\square$
\end{definition}

Finally, let us assume that the leaders $\mathcal{L}$ jointly optimize the system design. Once the system is designed, the followers react to this leader's strategic action by designing a selfish control input, i.e., the followers $\mathcal{M}$ play a dynamic game. Definition \ref{def:class_IV} shows he resulting Stackelberg game equilibrium for this combination of behavior across the layers (see Fig. \ref{fig:classes}).

\begin{definition}[Class IV: Stackelberg equilibrium with cooperative leaders and non-cooperative followers] 
\label{def:class_IV}
    A strategic profile $(\mathbf{a}^*,\mathbf{u}^*) \in \mathcal{A} \times \mathcal{U}$ is a Stackelberg equilibrium for cooperative leaders and non-cooperative followers if
        \begin{align*}
        &\tilde{\mathbf{u}}^* \in \mathbb{BR}^\mathcal{M}(\mathbf{a}^*),\\
        &\forall~j \in \mathcal{L}:a_j^* \in \arg\min_{a_j \in A_j} \hspace{-0.1cm}J_{j}(a_{j},a_{-j}^*,\tilde{\mathbf{u}}^*| \tilde{\mathbf{u}}^* \in \mathbb{BR}^\mathcal{M}(a_{j},a_{-j}^*)).
    \end{align*}
    In this case, all the leaders jointly design the parameters against which the followers will optimally respond by means of non-cooperative interactions. \hfill $\square$
\end{definition}

The Stackelberg equilibria for the different classes imply the computation of either a Nash equilibrium given by a fixed point condition or the solution of an optimization problem for the followers in terms of the leaders, and then the same procedure for the leader, knowing the optimal response of the followers against the leaders' decisions. In the following Section, algorithms to solve each of the classes are presented.

\vspace{-4mm}
\subsection{Learning Procedure to Compute Stackelberg Solutions}

This section provides a formal description of the procedures required to compute the Stackelberg equilibria for the four classes of games introduced in Fig.~\ref{fig:classes}. For each case, we present a qualitative overview of the computational steps along with the associated algorithmic implementation. We use the following notation: $|A_j|$ is the cardinality of the strategic set $A_j$ and $A_j(s)$ denotes the $s$-th element from $A_j$.

\subsubsection{Class I: Stackelberg game with non-cooperative leaders and non-cooperative followers}

Given that under this strategic behavior combination, both sets of decision-makers, i.e., leaders $\mathcal{L}$ and $\mathcal{M}$, play in a non-cooperative manner, the solution for the normal form game and dynamic game problems can be computed by following best-response dynamics. In this configuration, the leaders select their strategies independently and selfishly, anticipating the equilibrium response of the followers, who also act independently in optimizing their individual objectives.

The procedure for computing the Stackelberg equilibrium in this context involves the following steps:
\begin{itemize}
    \item  \textbf{Follower Response:} For each feasible leader strategic selection $\mathbf{a} \in \mathcal{A}$, let us compute the Nash equilibrium $\mathbf{u}^* \in \mathcal{U}$, which is denoted by $\mathbf{u}^*(\mathbf{a})$, meaning that the equilibrium strategic profile depends on the strategic selection of the leader. The computed Nash equilibrium is a non-cooperative best response against the leaders' strategies, i.e.,  $\mathbf{u}^*(\mathbf{a}) \in \mathbb{BR}^\mathrm{NC}_\mathcal{M}(\mathbf{a})$. 
    \item \textbf{Leader Game Solution:} Using the equilibrium strategic profile $\mathbf{u}^*(\mathbf{a})$ in terms of the leaders' selections, the leader non-cooperative game problem can be stated only in terms of $\mathbf{a}$, and the Nash equilibrium for the leaders $\mathbf{a}^*  \in \mathcal{A}$ can be found.
    \item \textbf{Follower Equilibrium Update:} Using the Nash equilibrium for the leaders game problem $\mathbf{a}^*$, the Nash equilibrium for the follower game problem becomes $\mathbf{u}^* := \mathbf{u}^*(\mathbf{a}^*)$.
    \item \textbf{Stackelberg Equilibrium:} One obtains the Class I Stackelberg equilibrium as $(\mathbf{u}^*,\mathbf{a}^*) \in \mathcal{U} \times \mathcal{A}$.
\end{itemize}

Details for this computation are presented in Algorithm \ref{alg:classI}.

\begin{algorithm}[t]
\caption{Learning Algorithm for Stackelberg equilibrium in Class I}\label{alg:classI}
\begin{algorithmic}
\State $\mathcal{L} = \{1,\dots,L\}$, $\mathcal{M} = \{1,\dots,M\}$ 
 $A_j,~\forall j \in \mathcal{L}$, 
 \State $j \gets 1,~i \gets 1,$ 
 $T$ 
\Procedure \text{Non-Cooperative Game for Followers}
\For{$s_1 \gets 1$ to $|A_1|$}
\State $\vdots$
\For{$s_L \gets 1$ to $|A_L|$}
    \State $a_1 \gets A_1(s_1)$,  $\dots$, $a_L \gets A_L(s_L)$ 
    \State $\mathbf{a} \gets (a_1,\dots,a_L) \in \mathcal{A}$ 
    \State $\mathbf{u} \in \mathcal{U}$ 
    \While{$\mathbf{u} = (\mathbf{u}_1,\dots,\mathbf{u}_M) \notin \epsilon-\mathbb{NE}_{\mathrm{follower}}$} 
        \For{$i \gets 1$ to $M$}
            \State $\mathbf{u}_i \gets \arg\min_{\mathbf{u}_i \in \mathcal{U}_i} V_i(x_0,\mathbf{u}_{i},\mathbf{u}_{-i})$ \State \text{s.~t.} \eqref{eq:dynamical_system}, $x_k \in \mathbb{X}(\mathbf{a}), \forall k\in[0..T]$
        \EndFor
    \EndWhile
    \State $\mathbf{u}^*(\mathbf{a}) \in \mathcal{U} \gets \mathbf{u}$ 
\EndFor
\EndFor
\EndProcedure
\Procedure \text{Non-Cooperative Game for Leaders}
\While{$\mathbf{a} = (a_1,\dots,a_L) \notin \epsilon-\mathbb{NE}_{\mathrm{leader}}$} 
    \For{$j \gets 1$ to $L$}
        \State $J_{j}(a_{j},a_{-j}) \gets J_{j}(a_{j},a_{-j},\mathbf{u}^*(\mathbf{a}))$ 
        \State $a_j \gets \arg\min_{a_j \in A_j} J_{j}(a_{j},a_{-j})$
    \EndFor
\EndWhile
\State $\mathbf{a}^* \gets \mathbf{a} \in \mathcal{A}$ 
\EndProcedure

\Procedure \text{Stackelberg Game Solution}
\State $\mathbf{u}^* \gets \mathbf{u}^*(\mathbf{a}^*) \in \mathcal{U}$ 
\State $(\mathbf{u}^*,\mathbf{a}^*) \in \mathcal{U} \times \mathcal{A}$ 
\EndProcedure
\end{algorithmic}
\end{algorithm}

\subsubsection{Class II: Stackelberg game with cooperative leaders and cooperative followers}

Unlike Class I, where both leaders and followers act independently in a non-cooperative manner, Class II assumes full cooperation within each layer. Here, leaders jointly coordinate their strategic decisions, and followers collaboratively optimize their control actions. This cooperative interaction transforms the hierarchical game into a sequence of centralized optimization problems rather than nested non-cooperative games.

The procedure to compute the Stackelberg equilibrium in this cooperative setting is as follows:
\begin{itemize}
    \item \textbf{Joint Follower Optimization:} For each feasible leader strategic selection $\mathbf{a} \in \mathcal{A}$, let us compute the optimal solution for the cooperative problem given by $\tilde{\mathbf{u}}^* \in \mathcal{U}$, which is denoted by $\tilde{\mathbf{u}}^*(\mathbf{a})$ as it depends on the strategic selection of the leader. The computed optimal is a cooperative best response against the leaders' strategies, i.e.,  $\tilde{\mathbf{u}}^*(\mathbf{a}) \in \mathbb{BR}^\mathrm{C}_\mathcal{M}(\mathbf{a})$.
    \item \textbf{Joint Leader Optimization:} Using the optimal strategic profile $\tilde{\mathbf{u}}^*(\mathbf{a})$ in terms of the leaders' selections, the leader cooperative game problem can be stated only in terms of $\mathbf{a}$, and the optimal solution for the leaders $\tilde{\mathbf{a}}^*  \in \mathcal{A}$ can be found.
    \item \textbf{Stackelberg Equilibrium:} Class II Stackelberg equilibrium can be found as $(\tilde{\mathbf{u}}^*,\tilde{\mathbf{a}}^*) \in \mathcal{U} \times \mathcal{A}$.
\end{itemize}

Class II replaces iterative best-response dynamics typical of non-cooperative games with a two-layer hierarchical optimization of coupled cost functions. This fundamental structural difference enables more tractable computations under cooperation assumptions, as outlined in Algorithm \ref{alg:classII}.

\begin{algorithm}[t]
\caption{Learning Algorithm for Stackelberg equilibrium in Class II}\label{alg:classII}
\begin{algorithmic}
\State $\mathcal{L} = \{1,\dots,L\}$, $\mathcal{M} = \{1,\dots,M\}$,  
 $A_j,~\forall j \in \mathcal{L}$ 
\State $j \gets 1,~i \gets 1,$ 
$T$ 
\Procedure \text{Control / Cooperative Game for Followers}
\For{$s_1 \gets 1$ to $|A_1|$}
\State $\vdots$
\For{$s_L \gets 1$ to $|A_L|$}
    \State $a_1 \gets A_1(s_1)$,  $\dots$, $a_L \gets A_L(s_L)$ 
    \State $\mathbf{a} \gets (a_1,\dots,a_L) \in \mathcal{A}$ 
            \State $\tilde{\mathbf{u}}^*(\mathbf{a}) \gets \arg\min_{\mathbf{u} \in \mathcal{U}} \sum_{i \in \mathcal{M}} V_i(x_0,\mathbf{u})$  
            \State s.~t  \eqref{eq:dynamical_system}, $x_k \in \mathbb{X}(\mathbf{a}), \forall k\in[0..T]$
    %
\EndFor
\EndFor
\EndProcedure
\Procedure \text{Control / Cooperative Game for Leaders}
        \State $J_{j}(\mathbf{a}) \gets J_{j}(\mathbf{a},\tilde{\mathbf{u}}^*(\mathbf{a}))$ 
        \State $\tilde{\mathbf{a}}^* \gets \arg\min_{\mathbf{a} \in \mathcal{A}} \sum_{j \in \mathcal{L}} J_{j}(\mathbf{a})$ 
\EndProcedure
\Procedure \text{Stackelberg Game Solution}
\State $\tilde{\mathbf{u}}^* \gets \tilde{\mathbf{u}}^*(\tilde{\mathbf{a}}^*) \in \mathcal{U}$ 
\State $(\tilde{\mathbf{u}}^*,\tilde{\mathbf{a}}^*) \in \mathcal{U} \times \mathcal{A}$ 
\EndProcedure
\end{algorithmic}
\end{algorithm}

\subsubsection{Class III: Stackelberg game with non-cooperative leaders and cooperative followers}

Class III represents a mixed strategic setting in which the leaders $\mathcal{L}$ and followers $\mathcal{M}$ adopt fundamentally different behaviors. Specifically, the leaders behave independently, each optimizing their own individual cost function without cooperation within the leader layer. This structure leads to a two-layer approach: a non-cooperative game among leaders solved via best-response dynamics, and a cooperative control problem solved collectively by the followers. The interplay between these layers defines the Stackelberg equilibrium for this class.

The computation steps of the Stackelberg equilibrium are analogous to the ones in Class I. First, for each feasible leader strategic selection $\mathbf{a} \in \mathcal{A}$, the followers solve a centralized cooperative optimal control problem to obtain their best response $\tilde{\mathbf{u}}^*(\mathbf{a}) \in \mathbb{BR}^\mathrm{C}_\mathcal{M}(\mathbf{a})$. Then, using this best response, the leaders engage in a non-cooperative game among themselves, where each optimizes its individual objective, leading to a Nash equilibrium $\mathbf{a}^*  \in \mathcal{A}$. The followers then update their strategy accordingly to $\tilde{\mathbf{u}}^* := \tilde{\mathbf{u}}^*(\mathbf{a}^*)$. Together, the pair $(\tilde{\mathbf{u}}^*,\mathbf{a}^*) \in \mathcal{U} \times \mathcal{A}$ constitutes the Stackelberg equilibrium for this class, capturing the strategic interaction between non-cooperative leaders and cooperative followers.

Class III captures realistic scenarios where leaders may compete or have conflicting interests, while followers coordinate actions for mutual benefit. Algorithm \ref{alg:classIII} provides a detailed procedural framework for computing this equilibrium.

\begin{algorithm}[t]
\caption{Learning Algorithm for Stackelberg equilibrium in Class III}\label{alg:classIII}
\begin{algorithmic}
\State $\mathcal{L} = \{1,\dots,L\}$, $\mathcal{M} = \{1,\dots,M\}$, 
$A_j,~\forall j \in \mathcal{L}$ 
\State $j \gets 1,~i \gets 1,$ 
$T$ 
\Procedure \text{Control / Cooperative Game for Followers}
\For{$s_1 \gets 1$ to $|A_1|$}
\State $\vdots$
\For{$s_L \gets 1$ to $|A_L|$}
    \State $a_1 \gets A_1(s_1)$,  $\dots$, $a_L \gets A_L(s_L)$ 
    \State $\mathbf{a} \gets (a_1,\dots,a_L) \in \mathcal{A}$ 
            \State $\tilde{\mathbf{u}}^*(\mathbf{a}) \gets \arg\min_{\mathbf{u} \in \mathcal{U}} \sum_{i \in \mathcal{M}} V_i(x_0,\mathbf{u})$ 
            \State s.~t. \eqref{eq:dynamical_system}, $x_k \in \mathbb{X}(\mathbf{a}), \forall k\in[0..T]$
    %
\EndFor
\EndFor
\EndProcedure
\Procedure \text{Non-Cooperative Game for Leaders}
\While{$\mathbf{a} = (a_1,\dots,a_L) \notin \epsilon-\mathbb{NE}_{\mathrm{leader}}$} 
    \For{$j \gets 1$ to $L$}
        \State $J_{j}(a_{j},a_{-j}) \gets J_{j}(a_{j},a_{-j},\tilde{\mathbf{u}}^*(\mathbf{a}))$ 
        \State $a_j \gets \arg\min_{a_j \in A_j} J_{j}(a_{j},a_{-j})$
    \EndFor
\EndWhile
\State $\mathbf{a}^* \gets \mathbf{a} \in \mathcal{A}$ 
\EndProcedure
\Procedure \text{Stackelberg Game Solution}
\State $\tilde{\mathbf{u}}^* \gets \mathbf{u}^*(\mathbf{a}^*) \in \mathcal{U}$ 
\State $(\tilde{\mathbf{u}}^*,\mathbf{a}^*) \in \mathcal{U} \times \mathcal{A}$ 
\EndProcedure
\end{algorithmic}
\end{algorithm}

\subsubsection{Class IV: Stackelberg game with cooperative leaders and non-cooperative followers}

Class IV  corresponds to the case where the leaders engage in cooperative behavior while the followers act in a non-cooperative, competitive manner. In this configuration, the leaders jointly solve an optimization problem to determine their collective optimal strategy, whereas the followers engage in a dynamic game, reaching equilibrium through best-response dynamics.

The computation of the Stackelberg equilibrium for this setting proceeds as follows. First, for every feasible leader strategic profile $\mathbf{a} \in \mathcal{A}$, the followers solve a non-cooperative dynamic game to compute the corresponding Nash equilibrium $\mathbf{u}^*(\mathbf{a}) \in \mathbb{BR}^\mathrm{NC}_\mathcal{M}(\mathbf{a})$, which represents the followers' best responses to the leaders’ decisions. Second, using the equilibrium strategic profile $\mathbf{u}^*(\mathbf{a})$ in terms of the leaders' selections, the leader cooperative problem can be stated only in terms of $\mathbf{a}$, and the joint optimization for the leaders can be computed, i.e., $\tilde{\mathbf{a}}^*  \in \mathcal{A}$ can be found. Then, using the cooperative solution for the leaders $\tilde{\mathbf{a}}^*$, the Nash equilibrium for the follower game problem becomes $\mathbf{u}^* := \mathbf{u}^*(\tilde{\mathbf{a}}^*)$. The resulting pair $(\mathbf{u}^*,\tilde{\mathbf{a}}^*) $ constitutes the Class IV Stackelberg equilibrium, integrating the leaders’ cooperative optimization with the followers’ strategic competition. The complete procedure for this computation is detailed in Algorithm~\ref{alg:classIV}.

\begin{algorithm}[t]
\caption{Learning Algorithm for Stackelberg equilibrium in Class IV}\label{alg:classIV}
\begin{algorithmic}
\State $\mathcal{L} = \{1,\dots,L\}$, $\mathcal{M} = \{1,\dots,M\}$, 
$A_j,~\forall j \in \mathcal{L}$ 
\State $j \gets 1,~i \gets 1,$ 
$T$ 
\Procedure \text{Non-Cooperative Game for Followers}
\For{$s_1 \gets 1$ to $|A_1|$}
\State $\vdots$
\For{$s_L \gets 1$ to $|A_L|$}
    \State $a_1 \gets A_1(s_1)$,  $\dots$, $a_L \gets A_L(s_L)$ 
    \State $\mathbf{a} \gets (a_1,\dots,a_L) \in \mathcal{A}$ 
    \State $\mathbf{u} \in \mathcal{U}$ 
    \While{$\mathbf{u} = (\mathbf{u}_1,\dots,\mathbf{u}_M) \notin \epsilon-\mathbb{NE}_{\mathrm{follower}}$} 
        \For{$i \gets 1$ to $M$}
            \State $\mathbf{u}_i \gets \arg\min_{\mathbf{u}_i \in \mathcal{U}_i} V_i(x_0,\mathbf{u}_{i},\mathbf{u}_{-i})$ \State s.~t.~ \eqref{eq:dynamical_system}, $x_k \in \mathbb{X}(\mathbf{a}), \forall k\in[0..T]$
        \EndFor
    \EndWhile
    \State $\mathbf{u}^*(\mathbf{a}) \in \mathcal{U} \gets \mathbf{u}$ 
\EndFor
\EndFor
\EndProcedure
\Procedure \text{Control / Cooperative Game for Leaders}
        \State $J_{j}(\mathbf{a}) \gets J_{j}(\mathbf{a},\mathbf{u}^*(\mathbf{a}))$ 
        \State $\tilde{\mathbf{a}}^* \gets \arg\min_{\mathbf{a} \in \mathcal{A}} \sum_{j \in \mathcal{L}} J_{j}(\mathbf{a})$ 
\EndProcedure
\Procedure \text{Stackelberg Game Solution}
\State $\mathbf{u}^* \gets \mathbf{u}^*(\tilde{\mathbf{a}}^*) \in \mathcal{U}$ 
\State $(\mathbf{u}^*,\tilde{\mathbf{a}}^*) \in \mathcal{U} \times \mathcal{A}$ 
\EndProcedure
\end{algorithmic}
\end{algorithm}

\vspace{-5mm}
\subsection{Computations for Four Classes}

While the four classes of Stackelberg games provide a structured and systematic framework for modeling heterogeneous strategic interactions, the computation of their corresponding equilibria can be computationally demanding, especially for large-scale systems with multiple decision-makers. The primary source of this computational burden stems from the necessity to solve either optimization problems or game-theoretic equilibrium problems. 

For Classes I and IV, the followers’ layer involves the computation of Nash equilibria in a dynamic game setting. This typically requires the implementation of iterative best-response algorithms or fixed-point methods, where each follower repeatedly adjusts their strategy based on the actions of others. The convergence of these iterative methods may require a substantial number of iterations, particularly when the interaction among players is highly nonlinear or when their strategy spaces are large. Moreover, because this Nash equilibrium computation must be repeated for every feasible strategic profile of the leaders, the computational cost scales quickly with the number of players and the size of their strategy sets.

For Classes II and III, while the followers act cooperatively and thus their problem reduces to solving a centralized optimal control problem, the leaders’ layer (Class III) or both layers (Class II) may still involve solving large-scale optimization problems over potentially high-dimensional spaces. Even though cooperative optimization problems tend to be more tractable than general Nash equilibrium computations, they can still be computationally heavy if the problem involves nonlinear dynamics, long planning horizons, or complex constraints.

In this paper, we consider finite feasible options for the leaders' strategic selections, which simplifies part of the computational process by enabling exhaustive search or combinatorial optimization techniques at the leader level. Additionally, the followers' best-response dynamics can be computed for each leader selection. 

\vspace{-5mm}
\subsection{Convergence Discussions of Algorithms}

The convergence of both best-response-based algorithms and optimization-based methods has been extensively studied in the literature, particularly for problems involving continuous, convex, and coercive cost functionals with respect to the decision variables (strategies), as we have stated in the respective problem statements, e.g., linear-quadratic difference games \cite{algorithm_nash,algorithm_3,algorithm_2,algorithm_1}. 

However, in the context of the Stackelberg game structures considered here, additional challenges arise. Specifically, the function $\mathbf{u}(\mathbf{a})$, representing the optimal response of the followers to a given leader strategy $\mathbf{a}$, introduces significant complexity. This makes the leader-level objective functions, $J_j(a_j, a_{-j}, \mathbf{u}(\mathbf{a}))$, $\forall j \in \mathcal{L}$, generally non-convex and difficult to handle analytically. Unlike standard Stackelberg game formulations where leaders and followers are coupled through shared dynamics or additive costs (e.g., \cite{BookBaTe, Jingrui2023, HUANG20202237}), here the leaders' strategies $\mathbf{a}$ affect the constraints of the follower optimization problem $\mathcal{P}_{\mathrm{follower}}$.

To address this challenge, the proposed algorithms compute a numerical approximation of the optimal follower response $\mathbf{u}(\mathbf{a})$ for each possible leader strategy $\mathbf{a}$. By working directly with these numerical solutions, the leader-level cost function $J_j(a_j, a_{-j}, \mathbf{u}(\mathbf{a}))$ can be structured to retain desirable properties, such as convexity with respect to $a_j$. This numerical decoupling simplifies the problem structure without requiring explicit expressions for $\mathbf{u}(\mathbf{a})$.

Provided that the follower cost functionals $V_i(x_0, \mathbf{u})$ are continuous, convex, and coercive with respect to each follower’s control $\mathbf{u}i$, and that the feasible set of $\mathcal{P}{\mathrm{follower}}$ is non-empty for each $\mathbf{a}$, existence of a solution to the follower subproblem is guaranteed. Similarly, given a numerically computed $\mathbf{u}^*$, the leader optimization problem also admits a solution under the assumed continuity, convexity, and coerciveness of $J_j$ with respect to each $a_j$.

\section{Networked System Application}
\label{sec:case_study}

In this section, we apply the proposed four classes in a real-world drinking water network. This case study demonstrates that the proposed framework can produce co-design solution for large-scale networked systems. Additionally, this framework can also be extended to other networked systems, such as power grids, transportation networks, and supply chains.

The Barcelona Drinking Water Network (DWN) in \cite[Chapter 3, Fig. 3.4]{book_thesis}, managed by the company \textit{Aguas de Barcelona (AGBAR)}, supplies drinking water to Barcelona city and its metropolitan area, utilizing water from the Ter and Llobregat rivers—regulated upstream by dams with a combined reservoir capacity of 600 hm³, alongside groundwater from the Bes\'{o}s River aquifer and supplementary wells. The network integrates four drinking water treatment plants: the Abrera and Sant Joan Desp\`{i} facilities (Llobregat River), the Cardedeu plant (Ter River), and the Bes\'{o}s plant (groundwater), with additional pumping infrastructure extracting from wells to achieve a total flow of approximately 7 m³/s. This is a well-known benchmark that has been used to illustrate networked control applications, game-theory-based control analysis, and optimization-based controllers. Here, we use this case study to evaluate and illustrate the Stackelberg game classes introduced in this paper.

Let us consider the Barcelona drinking water network as a networked multi-agent system to be studied for the Stackelberg game for the co-design problem. Let $f({x}_{k},{u}_{1,k},\dots,{u}_{M,k}) = f({x}_{k},\{{u}_{i,k}\}_{i \in \mathcal{M}})$ be a linear system as follows:
\begin{subequations}
\label{eq:water_system}
\begin{align}
{x}_{k+1} &= A {x}_{k} + \hspace{-0.1cm} \sum_{i \in \mathcal{M}} \hspace{-0.1cm} B_i {u}_{i,k} + B_l {d}_k,~\forall k \in [0..(T-1)],\\
x_0 &\in \mathbb{X}(\cdot)~\text{given},
\end{align}
\end{subequations}
where ${x} \in \mathbb{R}^{17}$ denotes the system states corresponding to the water level at each one of the reservoirs in the DWN, ${u} \in \mathbb{R}^{61}$ denotes the control inputs corresponding to the controllable flows determined by either valves or pumps throughout the DWN, and ${d} \in \mathbb{R}^{25}$ denotes the vector of time-varying demands, which are assumed to be obtained using forecasting methodologies. The feasible set for the system states is 
\begin{align*}
    \mathbb{X}(\cdot) &= \{{x} \in \mathbb{R}^{17}: {x}^\mathrm{min} \leq {x} \leq {x}^\mathrm{max}(\cdot)\},
\end{align*}
where $\mathbb{X}(\cdot) := \mathbb{X}(\{\text{designed parameter}\}) $ , $ {x}^\mathrm{max}(\cdot)  := {x}^\mathrm{max}(\{\text{designed parameter}\}) $. $\bar{{x}}^\mathrm{max} \in \mathbb{R}^{17}$ be a nominal parameter for the system state constraints, i.e., when there is no co-design problem under consideration, then ${x}^\mathrm{max} = \bar{{x}}^\mathrm{max}$.   

In addition, the feasible set for the control inputs is 
\begin{align*}
    \mathbb{U} = \{{u} \in \mathbb{R}^{61}: {u}^\mathrm{min} \leq {u} \leq {u}^\mathrm{max}\}.
\end{align*}

\subsection{Leaders and Followers in the Barcelona DWN}

In terms of the co-design problem of the Barcelona DWN, we consider $L=4$ leaders within the system in charge of the design of four reservoirs' dimensions, i.e.,  $\mathcal{L}=\{1,\dots,4\}$ corresponding to the states $x_{4}, x_{10}, x_{14}, x_{1}$, respectively (see Table \ref{tab:sub-systems_states}). Each leader in $\mathcal{L}$ has a set of possible actions in the strategic game $$A_j = \{0.5, 0.75, 1, 1.25, 1.5\},~ \forall~ j \in \mathcal{L}.$$ 
Also, we have that
$x_1^\mathrm{max} = a_1 \cdot \bar{x}_1^\mathrm{max}$,  
$x_4^\mathrm{max} = a_2 \cdot \bar{x}_4^\mathrm{max}$, 
$x_{10}^\mathrm{max} = a_3 \cdot \bar{x}_{10}^\mathrm{max}$, 
$x_{14}^\mathrm{max} = a_4 \cdot \bar{x}_{14}^\mathrm{max}$,     
where $\bar{x}_1^\mathrm{max}$, $\bar{x}_4^\mathrm{max}$, $\bar{x}_{10}^\mathrm{max}$, and $\bar{x}_{14}^\mathrm{max}$ are the nominal values for the reservoirs under design. For all the other system state maximum values, we have that $x_j^\mathrm{max}=\bar{x}_j^\mathrm{max}$ with $j \in \{1,\dots,17\}\setminus \{1,4,10,14\}$.

\begin{table}[t!]
    \centering
    \caption{Decision-makers $\{1,2,3,4\}$ corresponding to sub-systems.}
    \label{tab:sub-systems_states}
    \resizebox{\columnwidth}{!}{
    \begin{tabular}{cccc}
    \hline
    No. of Subsystem & Color   & Total Amount of States & Designed State \\
    \hline      
    1   & Green   & 4                      & $x_4$          \\
    2   & Blue    & 8                      & $x_{10}$       \\
    3   & Magenta & 3                      & $x_{14}$       \\
    4   & Red     & 2                      & $x_1$         \\
    \hline
    \end{tabular}
    }
\end{table}

\begin{figure}
    \centering
    \begin{tabular}{cc}
        \includegraphics[scale=0.25]{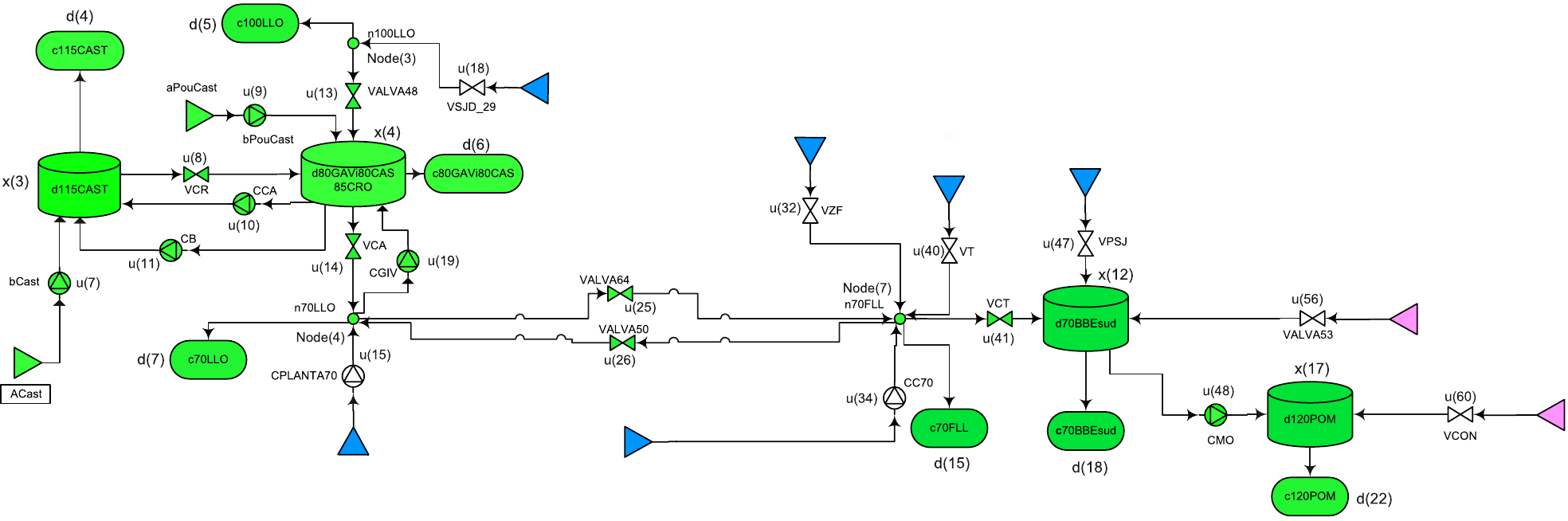} \\ 
        (a) \\
        \includegraphics[scale=0.25]{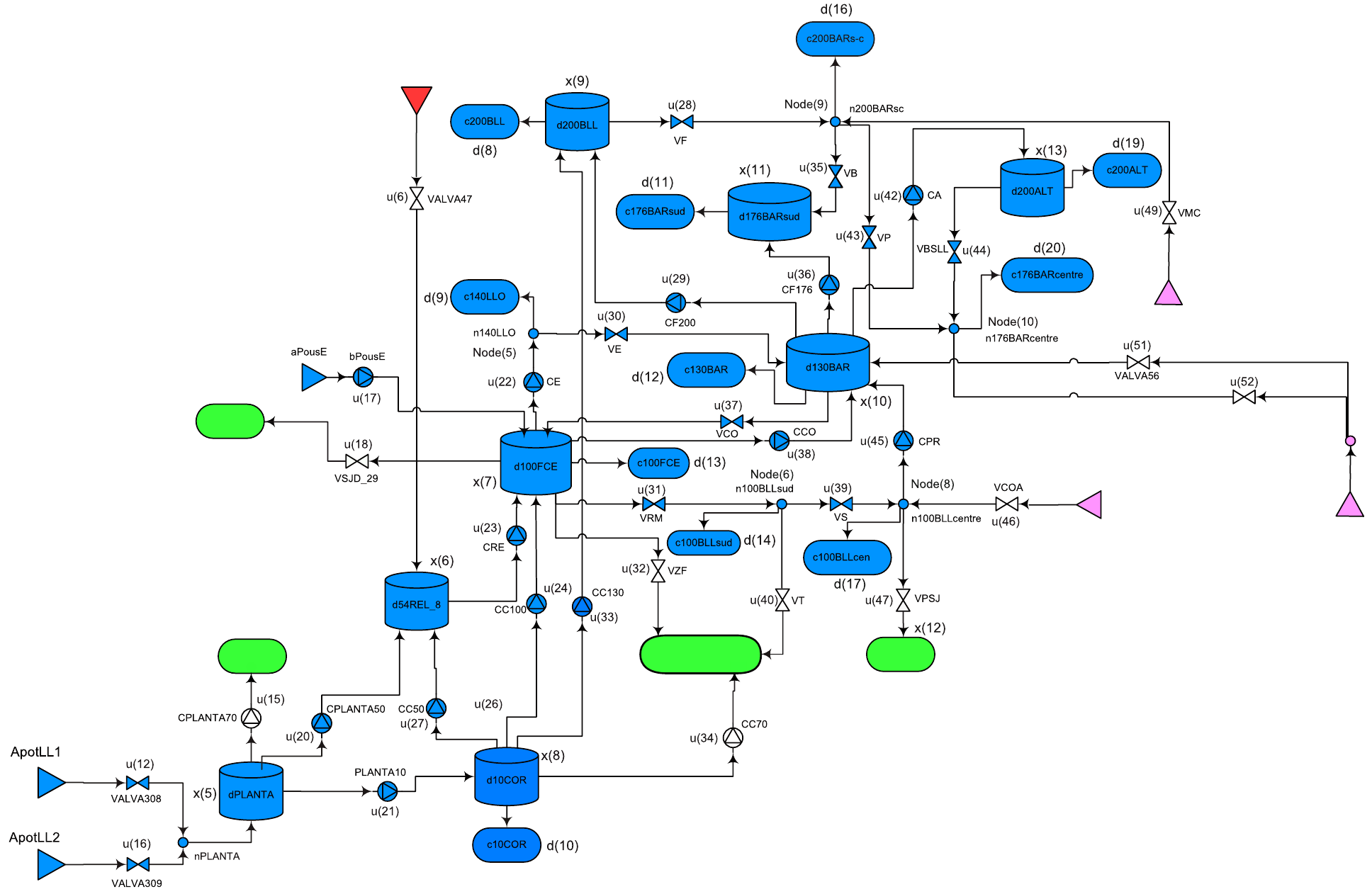} \\
         (b) \\ \\
         \begin{tabular}{cc}
              \includegraphics[scale=0.25]{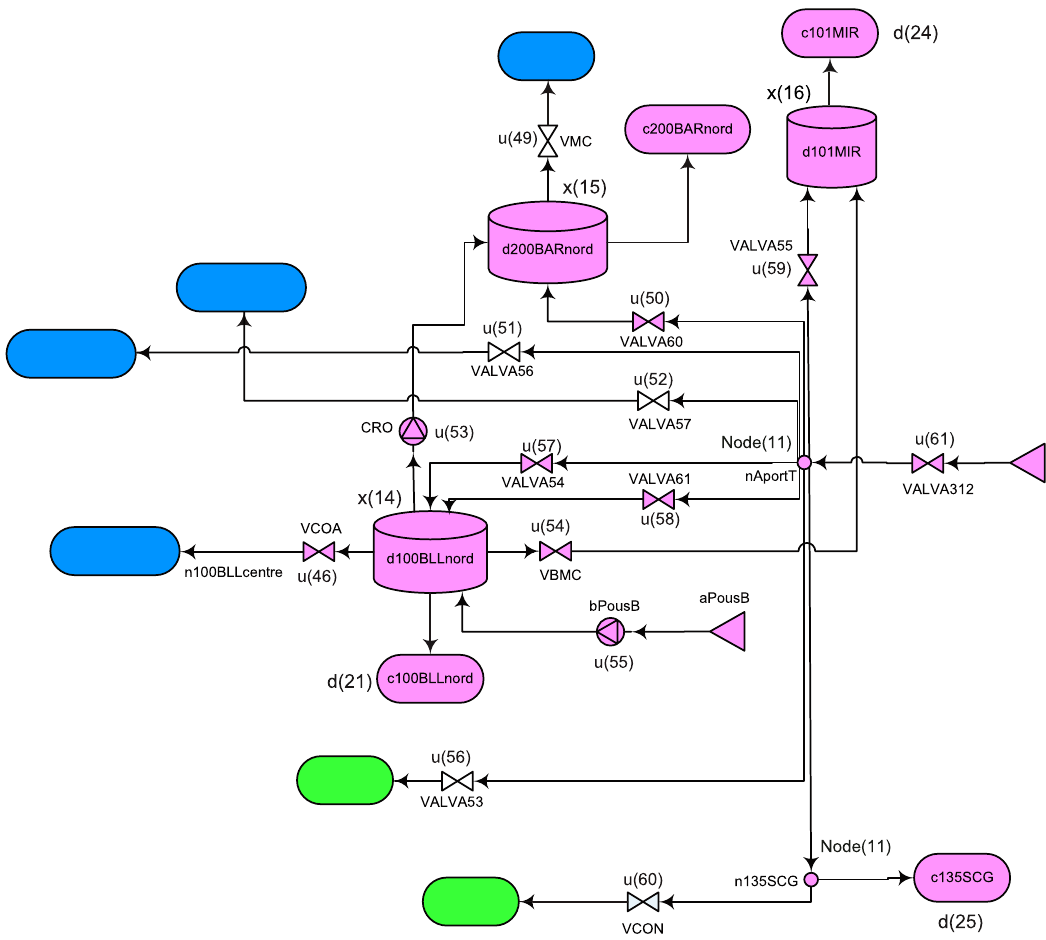} &
         \includegraphics[scale=0.25]{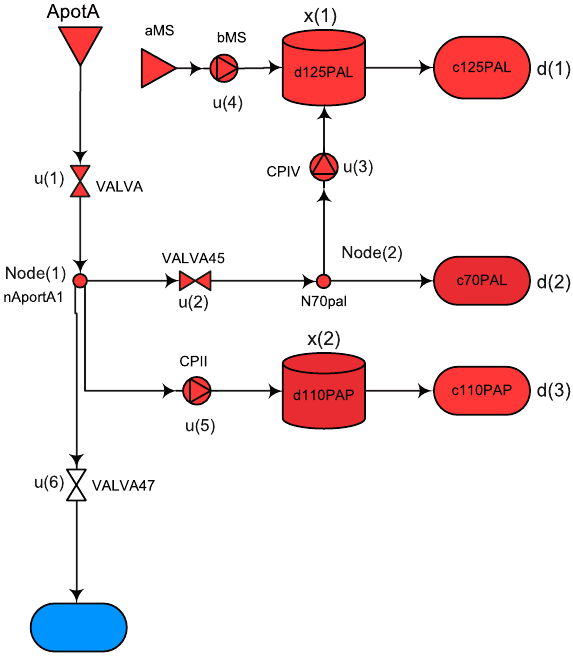}\\
          (c) & (d)
         \end{tabular}
    \end{tabular}
    \caption{Decision-makers in the followers' layer. (a) Decision-maker 1, (b) Decision-maker 2, (c) Decision-maker 3, (d) Decision-maker 4.}
    \label{fig:BCN Network players}
\end{figure}

On the other hand, let us consider $M=4$ followers, i.e., $\mathcal{M}=\{1,\dots,4\}$, which are in charge of four different sub-systems of the DWN. 
As shown in Fig. \ref{fig:BCN Network players}, four subsystems highlighted by different colors are considered. This partitioning has been adopted from the research reported in \cite[Chapter 8, Fig. 8.3]{book_thesis}. Thus, each follower decides the control inputs $\mathbf{u}_1,\dots,\mathbf{u}_4$ corresponding to the sub-systems with colors green, blue, magenta, and red, respectively. 

Note that the strategic selection of the followers $\mathcal{M}$ is influenced by the strategic selection of the leaders $\mathcal{L}$. When the leaders decide on a design for the reservoirs at each one of the sub-systems, the followers modify accordingly their optimal strategies. In the following, we introduce the game-theoretic problems for both the leaders and followers, considering either cooperative or non-cooperative behavior. Let us start by introducing the normal form game problems for the leaders.

\subsection{Non-Cooperative and Cooperative Leaders Game}

Let the leaders in $\mathcal{L}$ behave in a non-cooperative manner, i.e., each leader decides its strategies independently. The non-cooperative game problem is given by
%
\begin{align*}
    \forall~ j \in \mathcal{L}&:~~\min_{a_{j} \in A_{j}} J_{j}(a_{j},a_{-j},\mathbf{u}),
\end{align*}
with a leader cost functional of the form:
\begin{align*}
    J_{j}(a_{j},a_{-j},\mathbf{u}) &=  
    g(\mathbf{a}) + h(\mathbf{u}),~\forall~j \in \mathcal{L},
\end{align*}
where $g: \mathcal{A} \to \mathbb{R}$ and $h: \mathcal{U} \to \mathbb{R}$. For example, we may consider the following functions for the leader costs: 
\begin{align*}
     g(\mathbf{a}) &= \mathbf{a}^\top Q_j \mathbf{a} + v_j^\top \mathbf{a},\\   
     h(\mathbf{u}) &= \sum_{i \in \mathcal{M}} \sum_{k =0}^{T} \alpha_{i,k}^{\top} {u}_{i,k},
\end{align*}
where $ Q_j \succeq 0$ is weighting matrices, $v_j$ is a given vector. For leader costs, the parameters are chosen as  $  Q_j = \mathbf{0}$ (inspired by an economic cost shape), $v_j = 0.01 $, and $\alpha_{i,k}$ is the vector of time-varying electricity prices per input unit for follower $i$ at time $k$. The function $g(\mathbf{a})$ is used to penalize the effort that the leader applies in the design of the reservoirs. This cost can be associated with economic costs for implementing the design. In this regard, the leader is interested in minimizing the magnitude of its strategic selection. On the other hand, the cost function $h(\mathbf{u}$ depends on the followers' strategic selection. This means that, when the leader makes decisions over the modeling, it also takes into consideration how the followers will perform their control actions. 
Note that this is a game problem as the cost functional of the $j-$th decision-maker (leader) is affected by the decisions made by $\mathcal{L} \setminus \{j\}$ through the followers' actions $\mathbf{u}_i$, for all $i \in \mathcal{M}$. To emphasize this coupling, notice that the strategic design of a single leader, e.g., the $j-$th leader, affects the evolution of the control actions for all the followers $\mathcal{M}$ as they are dynamically coupled through \eqref{eq:dynamical_system}. Therefore, as all the leaders $\mathcal{L}$ take into consideration all the followers' actions in their cost functionals, the decisions of the $j-$th leader affects the cost of all the other leaders $\mathcal{L}\setminus \{j\}$ through $\mathbf{u}_i$ for all $i \in \mathcal{M}$.

Now, let us assume that all the leaders $\mathcal{L}$ agree on cooperating in the design of the system. Therefore, all the leaders jointly solve the following optimization problem:
\begin{align}
\min_{\mathbf{a} \in \mathcal{A}} \sum_{j \in \mathcal{L}} J_{j}(a_{j},a_{-j},\mathbf{u}).
\end{align}
This problem can be interpreted as a direct optimization of the design parameters, taking into consideration how the followers will react against the design. 

\subsection{Non-Cooperative and Cooperative Followers Game}

We next introduce the specific dynamic game problem for the followers in the DWN. Each follower decision-maker deciding over each sub-system of the DWN performs in a non-cooperative fashion. Then, the dynamic game problem for the DWN is as follows:
\begin{subequations}
    \begin{align*}
        &\forall i \in \mathcal{M}:\min_{\mathbf{u}_{i} \in \mathcal{U}_{i}} V_i(x_0,\mathbf{u}_{i},\mathbf{u}_{-i}), \\
        &\text{s.~t.}
        \begin{cases}
            {x}_{k+1} = A {x}_{k} + \sum_{i \in \mathcal{M}} B_i {u}_{i,k} + B_l {d}_k,\\
            0 = \sum\limits_{i \in \mathcal{M}} E_i {u}_{i,k} + E_d {d}_{k},\\
            {u}_{i,k} \in \mathbb{U}_i,~i \in \mathcal{M},\\
            {x}_{k} \in \mathbb{X}(\mathbf{a}),
        \end{cases}    
    \end{align*}
\end{subequations}
where the cost for each follower is given by
\begin{align}
    V_i(x_0,\mathbf{u})= \sum_{k=0}^{T}  \alpha_{i,k}^\top {u}_{i,k} + \Delta {u}_{i,k}^{\top} R_i \Delta {u}_{i,k},~ \forall i \in \mathcal{M},
\end{align}
where $\Delta {u}_{i,k} = {u}_{i,k} - {u}_{i,k-1} $ and $R_i \succ 0$, and $T> 0$ is a planning horizon. The feasible sets for both the control strategies and system states are as follows:
\begin{align}
    \mathbb{U}_i &:= \{{u}_i \in \mathbb{R}^{n_{u_i}} : {u}^{\mathrm{min}}_i \leq {u}_i \leq {u}^{\mathrm{max}}_i\},\\
    \mathbb{X}(\mathbf{a}) &:= \{{x} \in \mathbb{R}^{n_{x}} : {x}^{\mathrm{min}} \leq {x} \leq {x}^{\mathrm{max}}(\mathbf{a})\}.
\end{align}
It is important to highlight that the leaders' decisions directly affect the system state constraints for each one of the followers. In addition, note that modifying a single reservoir's constraint has an impact over the whole networked system, i.e., over the whole set of decision-makers $\mathcal{M}$ given the constraint given by the dynamical system in \eqref{eq:dynamical_system}.

If the followers decide to cooperate in order to define the appropriate control inputs, then the problem becomes a traditional optimal control problem. The cooperative game problem is as follows:
\begin{subequations}
\begin{align*}
&\min_{(\mathbf{u}_1,\dots,\mathbf{u}_M) \in \prod_{i \in \mathcal{M}} \mathcal{U}_i} \sum_{i \in \mathcal{M}} V_i(x_0,\mathbf{u}),\\
&\text{s.~t.}~
\begin{cases}
    {x}_{k+1} = A {x}_{k} +  B {u}_{k} + B_l {d}_k,\\
    0 = E {u}_{k} + E_d {d}_{k},\\
    {u}_{i,k} \in \mathbb{U}_i,~i \in \mathcal{M},\\
    {x}_{k} \in \mathbb{X}(\mathbf{a}),
\end{cases}
\end{align*}
\end{subequations}
where 
    $B = [B_1 \quad \dots \quad B_M]$, 
    $E = [E_1 \quad \dots \quad E_M]$,  and
    ${u}_{k} = [{u}_{1,k}^\top \quad \dots \quad {u}_{M,k}^\top]^\top$. 
    %
We compute and test each one of the Stackelberg game classes presented in Fig. \ref{fig:classes} by combining the aforementioned game problems. The results are presented and discussed in the coming section, where we present the Stackelberg equilibrium for each class and we also analyze the price of anarchy at each layer (leader and follower layer).

\section{Results and Discussions}
\label{sec:results}

\begin{figure*}[t!]
\centering
    \subfigure[Subsystem 1]{\includegraphics[width=0.45\hsize]{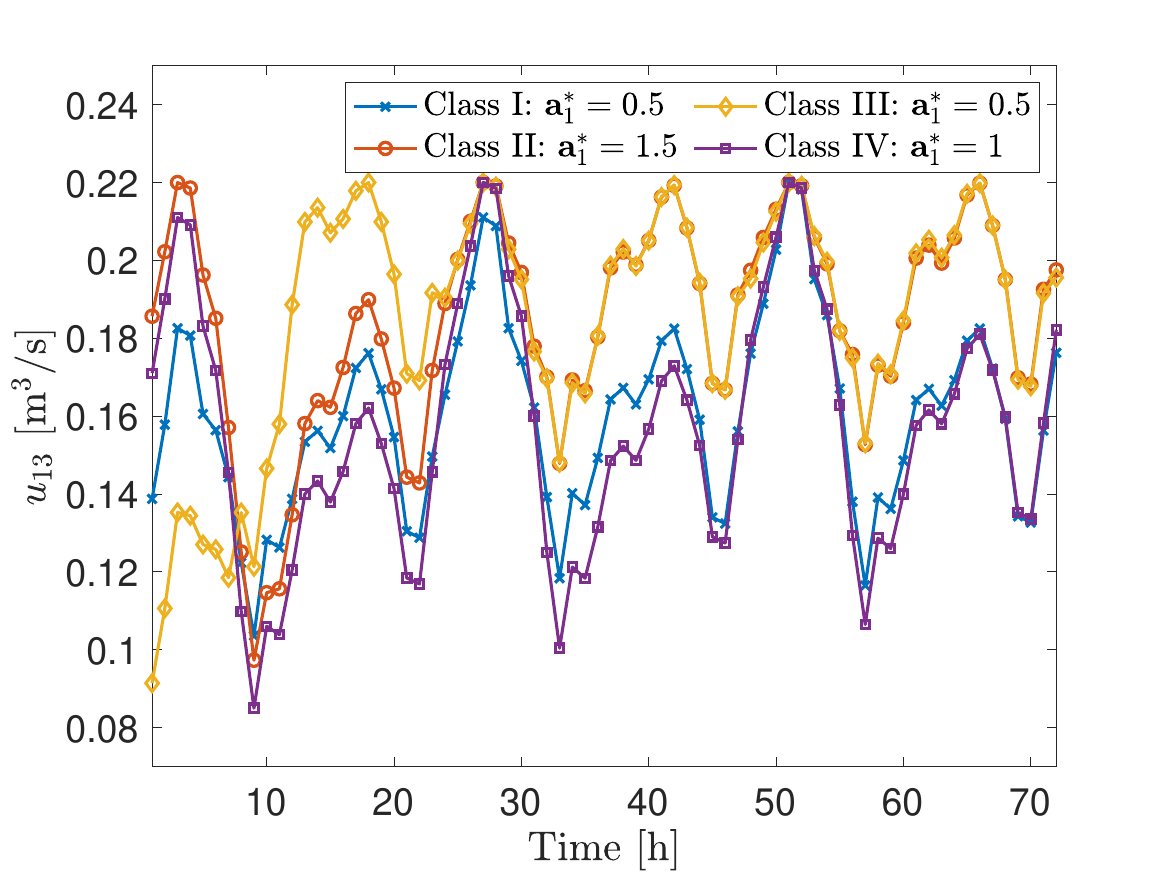}\label{fig:u-s1}}
    \subfigure[Subsystem 2]{\includegraphics[width=0.45\hsize]{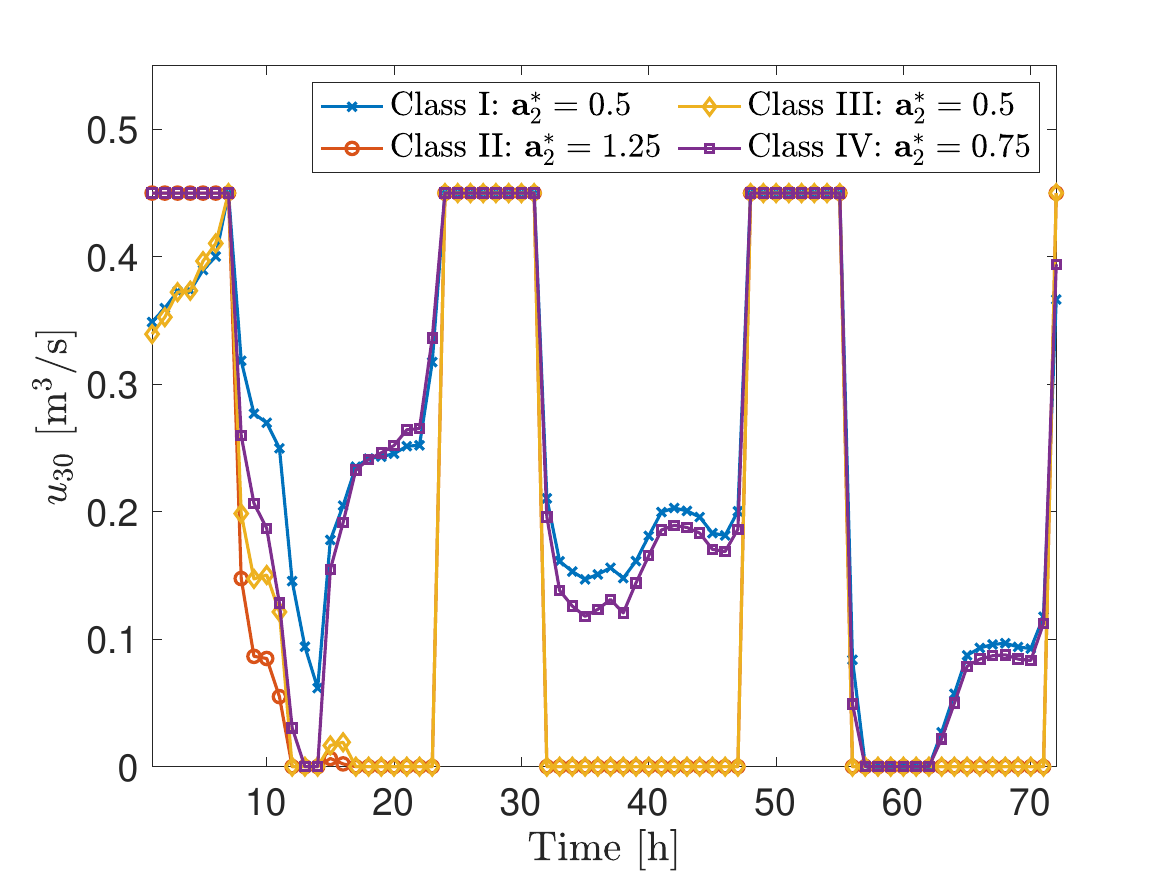}\label{fig:u-s2}}
    \subfigure[Subsystem 3]{\includegraphics[width=0.45\hsize]{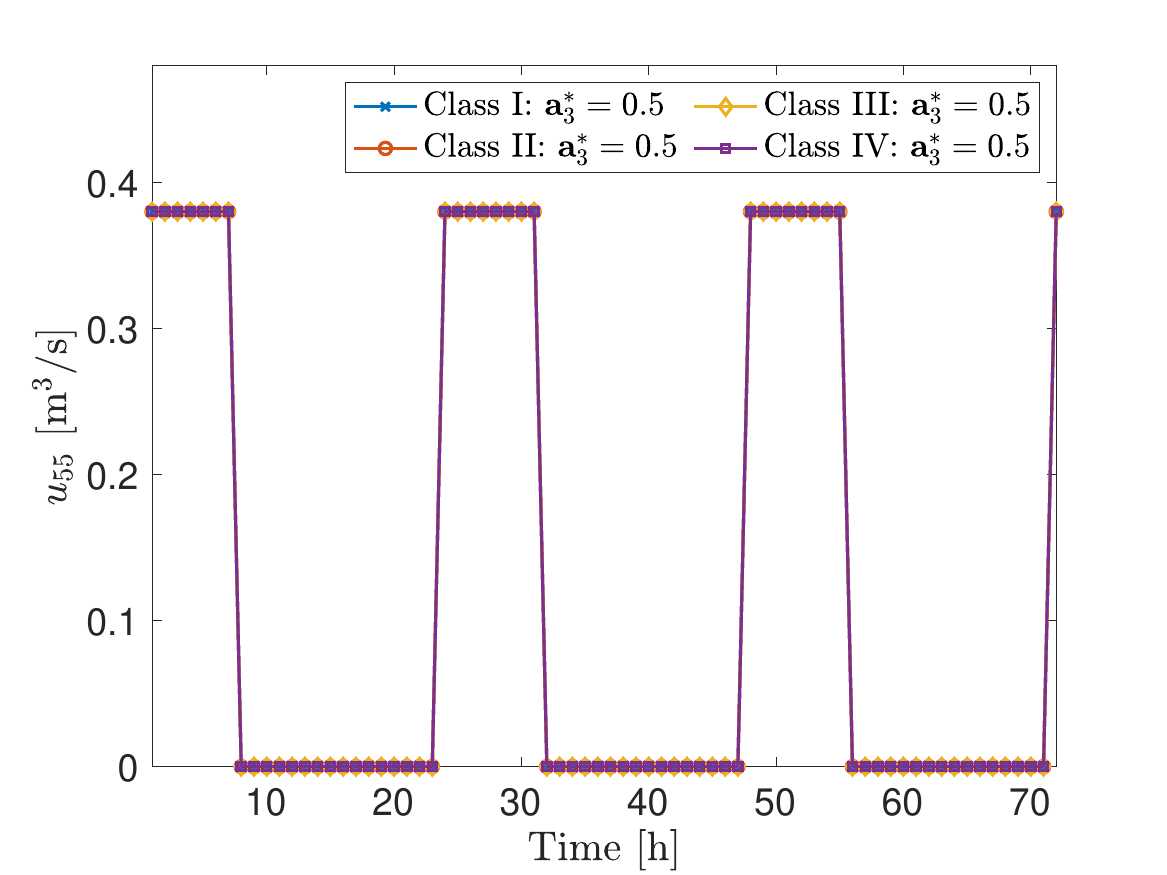}\label{fig:u-s3}}
    \subfigure[Subsystem 4]{\includegraphics[width=0.45\hsize]{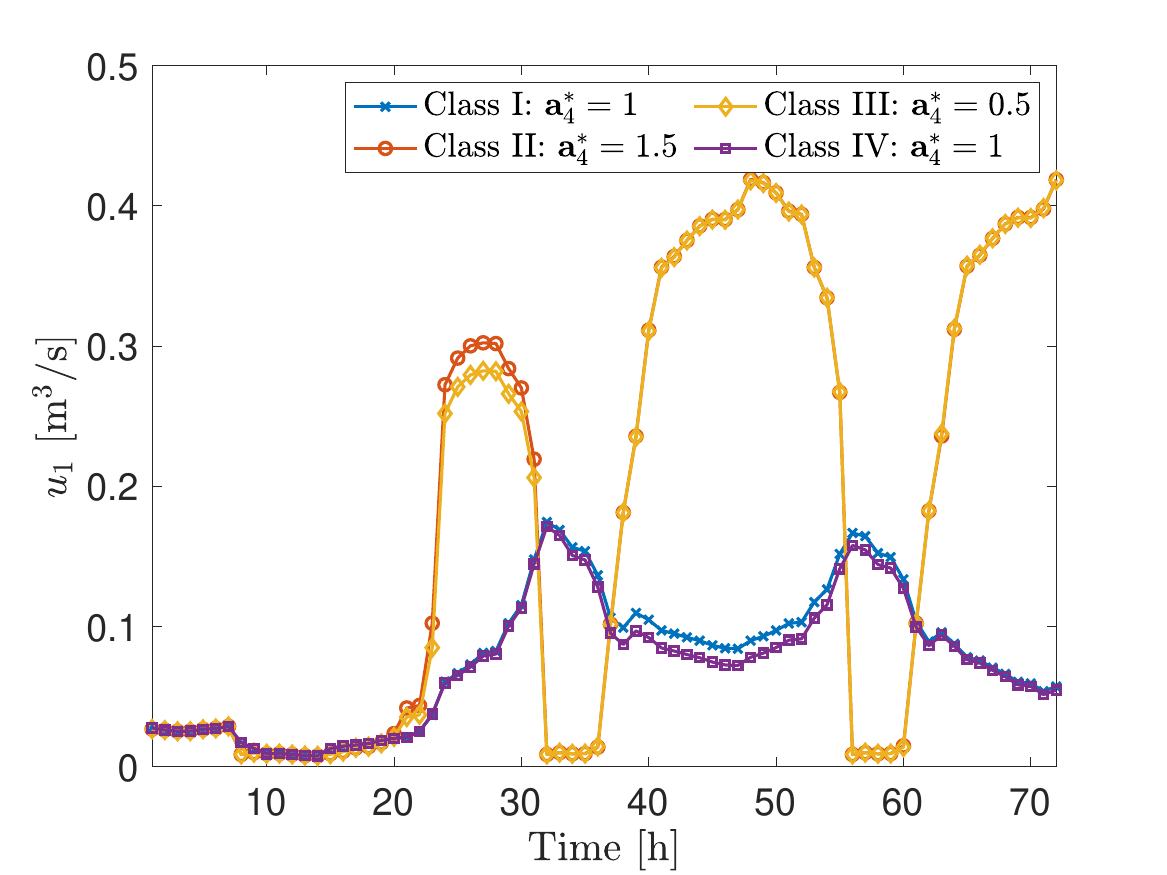}\label{fig:u-s4}}
    \caption{Optimal follower solutions with four classes.} %
    \label{fig:u}
\end{figure*}

\begin{figure*}[t!]
\centering
    \subfigure[Subsystem 1]{\includegraphics[width=0.45\hsize]{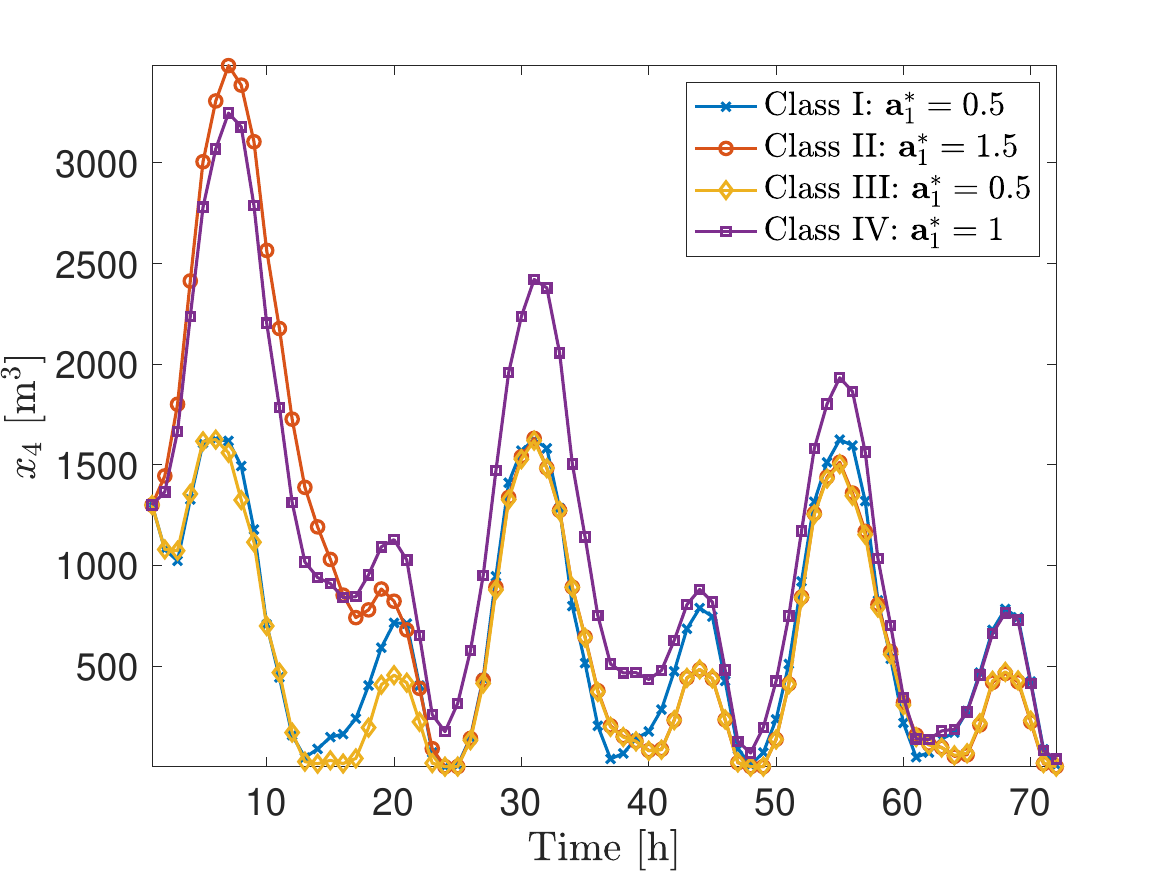}\label{fig:x-s1}}
    \subfigure[Subsystem 2]{\includegraphics[width=0.45\hsize]{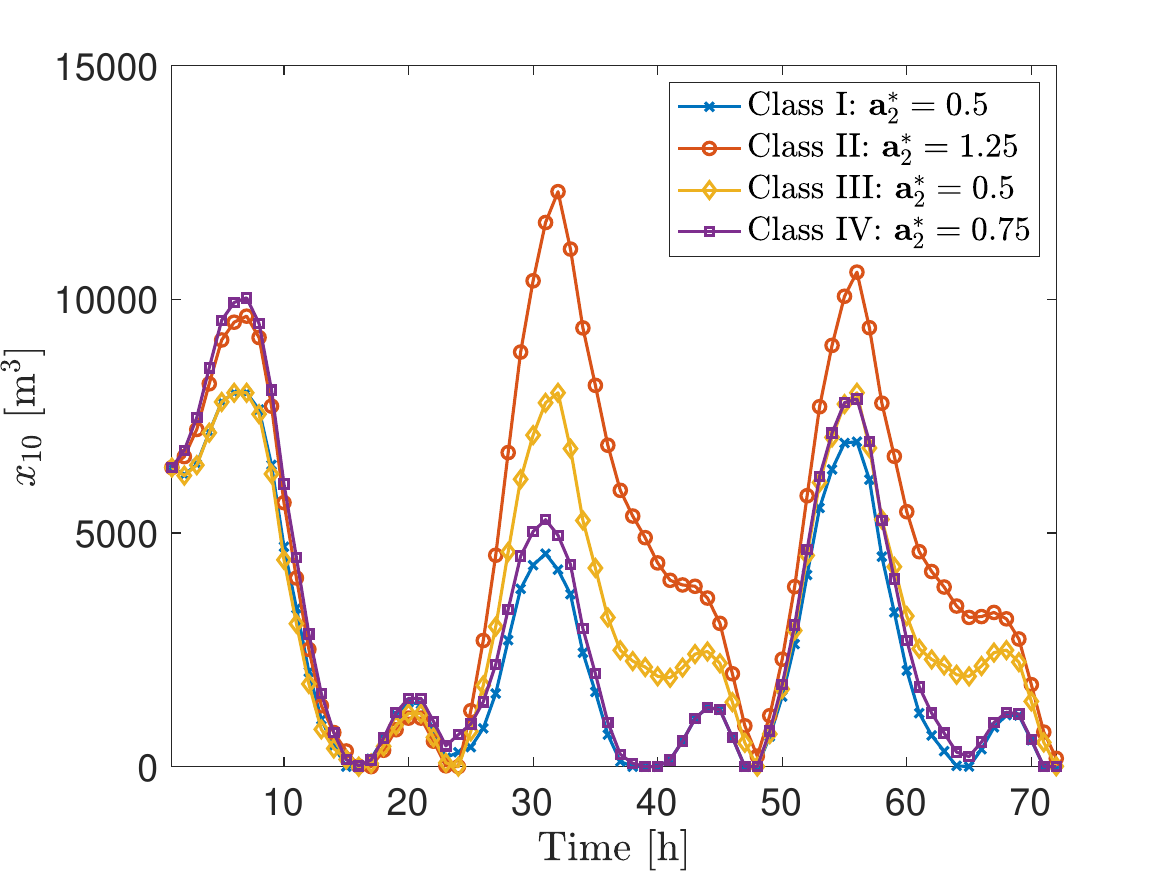}\label{fig:x-s2}}
    \subfigure[Subsystem 3]{\includegraphics[width=0.45\hsize]{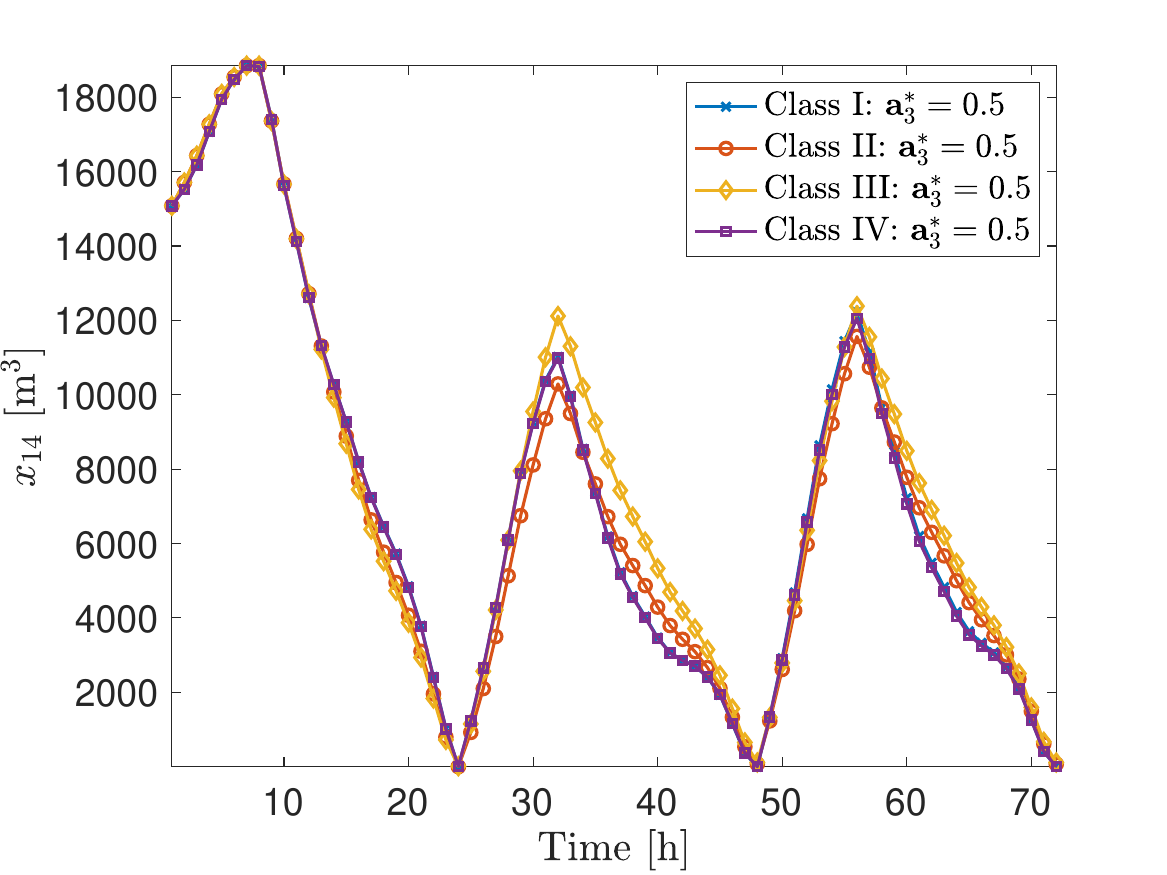}\label{fig:x-s3}}
    \subfigure[Subsystem 4]{\includegraphics[width=0.45\hsize]{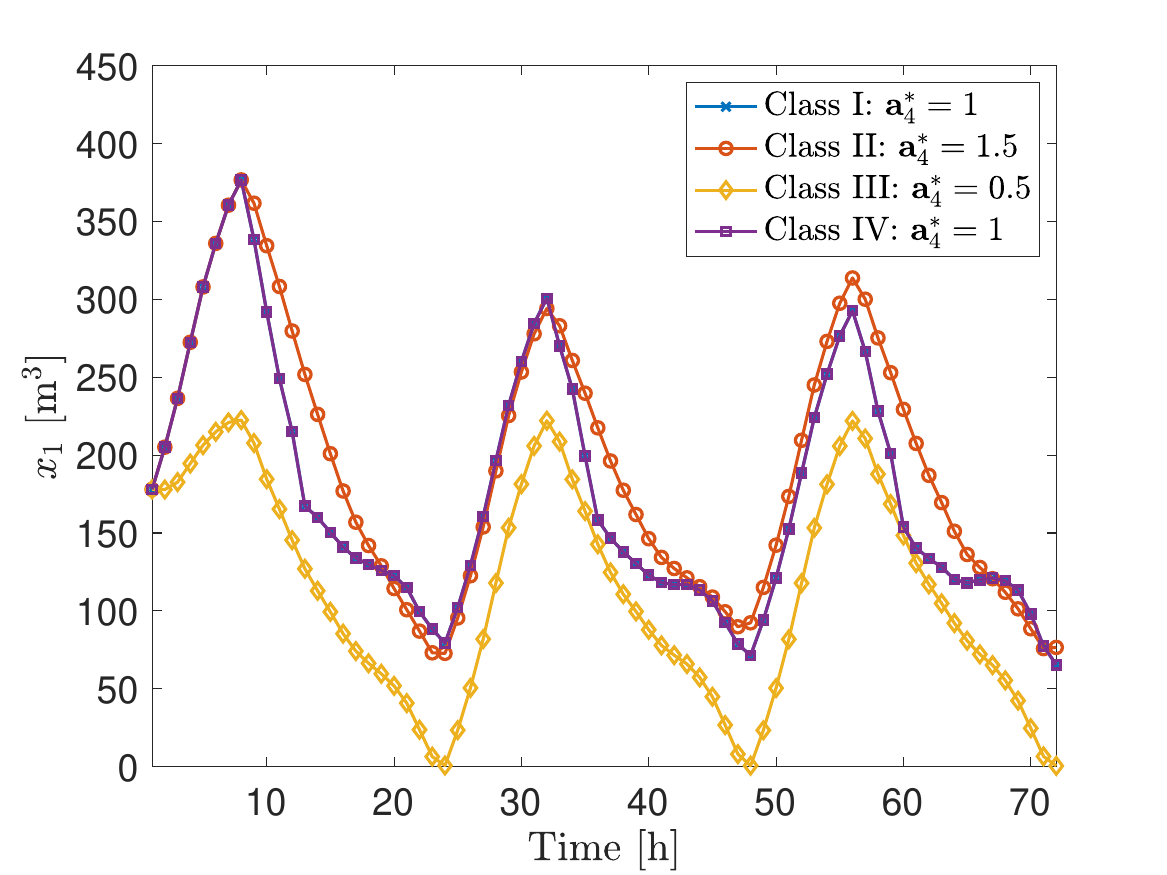}\label{fig:x-s4}}
    \caption{State evaluations with optimal leader and follower solutions.}
    \label{fig:x}
\end{figure*}

\begin{table*}[h]
    \centering
    \caption{Optimal leader solutions with four classes. %
    }
    \label{tab:equ_stackelberg_1}
    \begin{tabular}{lcccccccc}
    \hline
              & Leader          & Follower     & $\mathbf{a}^*$ & $J_1^*$ & $J_2^*$ & $J_3^*$ & $J_4^*$ & $J^*$ \\
    \hline
    Class I   & Non-cooperative & Non-cooperative &  $[0.5, \; 0.5,\;0.5,\; 1]$     
     & 252,647 & 252,711 & 252,819 & 252,635         &  1,010,812     \\
    Class II  & Cooperative     & Cooperative     &  $[1.5, \; 1.25,\;0.5,\; 1.5]$    & 250,340 & 250,491 & 250,479 &  250,297      &  1,001,607      \\
    Class III & Non-cooperative & Cooperative     &  $[0.5, \; 0.5,\;0.5,\; 0.5]$    
     & 250,386 & 250,449 & 250,558 & 250,372         & 1,001,765     \\
    Class IV  & Cooperative     & Non-cooperative &  $[1, \; 0.75,\;0.5,\; 1]$     & 252,620 & 252,707 & 252,776 & 252,592    &   1,010,695    \\
    \hline
    \end{tabular}
    \vspace{-0.2cm}
\end{table*}

\begin{table*}[h]
    \centering
    \caption{Optimal follower solutions with four classes. %
    }
    \label{tab:equ_stackelberg_2}
    \begin{tabular}{lcccccccc}
    \hline
              & Leader          & Follower     & $\mathbf{a}^*$ & $V_1^*$ & $V_2^*$ & $V_3^*$ & $V_4^*$ & $V^*$ \\
    \hline
    Class I   & Non-cooperative & Non-cooperative &  $[0.5, \; 0.5,\;0.5,\; 1]$     
     & 0.0732  & 17.4186  & 18.3429  &  0.7798 & 36.6145 \\
    Class II  & Cooperative     & Cooperative     &  $[1.5, \; 1.25,\;0.5,\; 1.5]$   & 0.0743  & 17.0877 &  17.5127  &  1.7343 & 36.4090\\
    Class III & Non-cooperative & Cooperative     &  $[0.5, \; 0.5,\;0.5,\; 0.5]$    
& 0.0740 &  17.0852  & 17.5375  &  1.7099 &  36.4066 \\
    Class IV  & Cooperative     & Non-cooperative &  $[1, \; 0.75,\;0.5,\; 1]$    & 0.0716 &  17.3884  & 18.3939  &  0.7383 &  36.5923\\
    \hline
    \end{tabular}
\end{table*}

\begin{table}[th]
    \centering
    \caption{Price of Anarchy computation results.}
    \label{tab:PoA}
    \begin{tabular}{llc}
    \hline
                              & \textbf{Scenario}                           & $\mathrm{PoA}$ \\
    \hline
    \multirow{2}{*}{Leader}   & Cooperative followers          &  1.0002   \\
                              & Non-cooperative followers   &  1.0001   \\
    \hline
    \multirow{2}{*}{Follower} & Cooperative leaders           &  1.0050   \\
                              & Non-cooperative leaders    &  1.0057   \\
    \hline
    \end{tabular}
\end{table}

All simulations for followers were conducted over a 72-hour period (3 days) with a sampling time interval of 1 hour. Water demands and electricity prices exhibit distinct daily patterns. The selected optimal solutions of the followers are shown in Fig.~\ref{fig:u}. In Figs.~\ref{fig:u-s1} and~\ref{fig:u-s2}, the optimal follower solutions vary significantly, primarily attributable to differences in tank designs across four distinct classes and the implementation of two different control strategies. Specifically, Fig. \ref{fig:u-s2} highlights the scenarios where the inputs occasionally reach the maximum values because identical input constraints are applied. In Fig. \ref{fig:u-s3}, due to the same leader solutions with four classes, the same optimal follower solutions are observed. Fig. \ref{fig:u-s4} further shows the impact of two control strategies: the follower solutions with Class I and Class IV, both governed by a non-cooperative control framework, exhibit striking similarities, as do those for Classes II and III, which adopt a cooperative control approach. Moreover, it can also be observed that the follower solutions, i.e., the flows through actuators (valves and pumps), have a potential daily pattern due to the water demand satisfaction.

Fig.~\ref{fig:x} illustrates the volume evolutions of selected tanks across four subsystems, each corresponding to one of the four classes. The observed daily pattern in tank volumes mirrors the water demand cycle. In Fig. \ref{fig:x-s3}, the volume evolutions of tank $x_{14}$ are similar to the four classes, as the designed tank sizes are the same based on the leader solutions, which is consistent with the follower solutions shown in Fig. \ref{fig:u-s3}.
From the game theory perspective, the trajectories presented in Fig. \ref{fig:u-s1} and Fig. \ref{fig:u-s4} correspond to the Nash equilibrium for the followers when they strategically interact in a dynamic game (Classes I and IV). In contrast, the trajectories in Fig. \ref{fig:u-s2} and Fig. \ref{fig:u-s3} present the optimal control inputs corresponding to a cooperative dynamic game (Classes II and III). The reader may compare the followers' strategic interactions in Fig. \ref{fig:classes}. 

Regarding the strategic selection for the leaders, this is presented in Tables \ref{tab:equ_stackelberg_1} and \ref{tab:equ_stackelberg_2}. The Nash equilibrium for the leaders correspond to $\mathbf{a}^* = [0.5, \; 0.5,\;0.5,\; 1]$ and $\mathbf{a}^* = [0.5, \; 0.5,\;0.5,\; 0.5]$ for the Class I and Class III, respectively. Interestingly, we see many similarities between both strategic profiles. Note that, when changing the strategic behavior of the followers from non-cooperative to cooperative, only one of the leaders deviates from its strategic selection. For the cooperative scenario for the leaders, we observe that the optimal solutions are $\mathbf{a}* = [1.5, \; 1.25,\;0.5,\; 1.5]$ and $\mathbf{a}* = [1, \; 0.75,\;0.5,\; 1]$ exhibiting a strategic deviation for all the leaders, except one, when the followers change their behavior from non-cooperative to cooperative. 

The optimal costs for both leaders and followers, and for all the possible classes (from Class I to Class IV), are presented in Table \ref{tab:equ_stackelberg_1} and Table \ref{tab:equ_stackelberg_2}. By using such optimal values, one can measure or evaluate the difference between cooperating and non-cooperating, or from the control perspective, one can evaluate the cost difference of a centralized controller in front of a decentralized controller. We perform this assessment by means of the price-of-anarchy introduced in Section \ref{sec:problem_statement}. The results of the price-of-anarchy corresponding to all the interactive combinations are presented in Table \ref{tab:PoA}. It is interesting to observe that the price-of-anarchy is quite close to one in all the cases, indicating that the obtained Nash equilibria are optimal.%

\section{Concluding Remarks and Future Directions}
\label{sec:conclusions}

We have presented multiple classes of Stackelberg games for the co-design of networked systems comprising the simultaneous design of both system and control. Under this approach, a leader is in charge of deciding on a design system parameter. Note that, in general, this decision can be related to the selection of elements such as actuators, or any other system specification. Then, there is a follower layer where the control design takes place. As shown in this paper, the decisions made at the control design depend on the decisions made at the system design stage. Hence, we have presented multiple possibilities for such a bi-level Stackelberg-like interaction. This is because we can consider multiple parties at each one of the layers, i.e., multiple leaders and multiple followers, leading to more involved game-theoretical settings. We consider the case in which leaders and followers can either cooperate or not, and all the possible combinations for these interactions. We have shown that the cooperative game approach coincides with a control problem, and the non-cooperative game can be seen as a decentralized control strategy. 
Moreover, the evaluation of the price-of-anarchy for the computed equilibrium solutions shows that the decentralized controllers are optimal. Finally, as it was highlighted in the manuscript, we have considered a discrete finite set of strategies for the leaders. 

As future work, we propose extending the framework to consider a continuous strategy space for the leaders. This extension would leverage approximation theory and learning-based methods to address the computational tractability challenges associated with large-scale systems. By adopting these techniques, it becomes feasible to approximate the optimal strategies in high-dimensional settings where exhaustive enumeration is intractable.

\bibliographystyle{unsrt}
\bibliography{references.bib}

\end{document}